\documentclass[3p]{elsarticle}

\usepackage{stfloats}
\usepackage{cuted}
\usepackage{caption}

\usepackage{indentfirst}
\usepackage[font=small, labelfont=bf]{caption}
\usepackage[labelformat=empty, position=top]{subcaption}
\usepackage[export]{adjustbox}
\usepackage{lineno}

\usepackage{graphicx}
\usepackage{dcolumn}
\usepackage{bm}

\usepackage{hyperref}
\usepackage{xr} 

\usepackage{rotating}
\usepackage{xspace}
\usepackage{tikz}
\usepackage{threeparttable}
\usepackage{booktabs}
\usepackage{comment}
\usepackage[utf8]{inputenc}
\usepackage[T1]{fontenc}

\usepackage{amsmath}
\usepackage{xcolor}
\usepackage{color,soul}
\usepackage{siunitx}

\usepackage[normalem]{ulem}
\RequirePackage[normalem]{ulem} 
\RequirePackage{color}\definecolor{RED}{rgb}{1,0,0}\definecolor{BLUE}{rgb}{0,0,1} 

\graphicspath{{./}}

\newcommand{\rev}[2]{\textcolor{orange}{[\textbf{REV #1}: #2]}}

\newcommand{\be}{\begin{enumerate}}
\newcommand{\ee}{\end{enumerate}}
\newcommand{\bi}{\begin{itemize}}
\newcommand{\ei}{\end{itemize}}

\newcounter{saveenumi}

\newcommand{\eq}[1]{Eq.~\ref{eq:#1}}

\newcommand{\fig}[1]{Fig.~\ref{fig:#1}}
\newcommand{\figSS}[1]{Fig.~S\ref{fig:#1}}
\newcommand{\figs}[1]{Figs.~\ref{fig:#1}}
\newcommand{\sect}[1]{Methods section~\ref{sec:#1}}

\newcommand{\tablSS}[1]{Table S\ref{table:#1}}

\newcommand{\etal}{{\it et al.~}}

\newcommand{\nba}[1]{}
\newcommand{\mAA}{{\text \AA}\xspace}
\newcommand{\nm}{nm\xspace}       

\newcommand{\horizline}{\begin{center}\line(1,0){500}\end{center}}

\newcommand{\thebmg}{~$\mathrm{Zr_{65}Cu_{17.5}Ni_{10}Al_{7.5}}$~}

\renewcommand{\rev}[2]{#2}  

\makeatletter
\newcommand*{\addFileDependency}[1]{
  \typeout{(#1)}
  \@addtofilelist{#1}
  \IfFileExists{#1}{}{\typeout{No file #1.}}
}
\makeatother

\newcommand*{\myexternaldocument}[1]{%
    \externaldocument{#1}%
    \addFileDependency{#1.tex}%
    \addFileDependency{#1.aux}%
}

\myexternaldocument{20yr_snem_SI}

\newcommand{\mytitle}{Mapping Structural Heterogeneity at the Nanoscale with \\ Scanning Nano-structure Electron Microscopy (SNEM)}

\newcommand{\addapam}{Department of Applied Physics and Applied Mathematics, Columbia University, New York, NY 10027, United States}
\newcommand{\addjhu}{Department of Materials Science and Engineering, Johns Hopkins University, Baltimore, Maryland 21218, United States}
\newcommand{\adducrmat}{Materials Science and Engineering Program, University of California, Riverside, Riverside CA 29521, United States}
\newcommand{\adducrmeche}{Department of Mechanical Engineering, University of California, Riverside, Riverside, CA, 92521, United States}
\newcommand{\addnm}{NanoMEGAS SPRL, Rue Émile Claus 49 Bte 9, B1050, Brussels, Belgium}
\newcommand{\addbnl}{Condensed Matter Physics and Materials Science Department, Brookhaven National Laboratory, Upton, NY 11973}

\begin{document}

\title{\mytitle}

\address[1]{\addapam}
\address[2]{\addjhu}
\address[3]{\addnm}
\address[4]{\adducrmat}
\address[5]{\adducrmeche}
\address[6]{\addbnl}

\author[1,2]{Yevgeny Rakita \corref{corr}}
\ead{yr2369@columbia.edu}

\author[2]{James~L. Hart \corref{eqcon}}

\author[3]{Partha Pratim Das \corref{eqcon}}

\author[4]{Sina Shahrezaei}

\author[2]{Daniel~L. Foley}

\author[4,5]{Suveen Nigel Mathaudhu}

\author[3]{Stavros Nicolopoulos}

\author[2]{Mitra~L. Taheri \corref{corr}}
\ead{mtaheri4@jhu.edu}

\author[1,6]{Simon~J.~L. Billinge \corref{corr}}
\ead{sb2896@columbia.edu}

\cortext[corr]{Corresponding authors}
\cortext[eqcon]{These authors contributed equally}

\date{\today}

\begin{abstract}
Here we explore the use of scanning electron diffraction (also known as 4D-STEM) coupled with electron atomic pair distribution function analysis (ePDF) to understand the local order (structure and chemistry) as a function of position in a complex multicomponent system, a hot rolled, Ni-encapsulated, Zr$_{65}$Cu$_{17.5}$Ni$_{10}$Al$_{7.5}$ bulk metallic glass (BMG), with a spatial resolution of 3~nm.  We show that it is possible to gain insight into the chemistry and chemical clustering/ordering tendency in different regions of the sample, including in the vicinity of nano-scale crystallites that are identified from virtual dark field images and in heavily deformed regions at the edge of the BMG. In addition to simpler analysis, unsupervised machine learning was used to extract partial PDFs from the material, modelled as a quasi-binary alloy, and map them in space. These maps allowed key insights not only into the local average composition, as validated by EELS, but also a unique insight into chemical short-range ordering tendencies in different regions of the sample during formation. The experiments are straightforward and rapid and, unlike spectroscopic measurements, don't require energy filters on the instrument. We spatially map different quantities of interest (QoI's), defined as scalars that can be computed directly from positions and widths of ePDF peaks or parameters refined from fits to the patterns. We developed a flexible and rapid data reduction and analysis software framework that allows experimenters to rapidly explore images of the sample on the basis of different QoI's. The power and flexibility of this approach are explored and described in detail. Because of the fact that we are getting spatially resolved images of the nanoscale structure obtained from ePDFs we call this approach scanning nano-structure electron microscopy (SNEM), and we believe that it will be powerful and useful extension of current 4D-STEM methods.
\label{sec:abstract}
\end{abstract}



\date{\today}

\maketitle



\section{Introduction}
Properties of complex materials depend on the spatial nature of ordering, which therefore must be well characterized.
Structural order can form on multiple length scales and this requires tools that can yield local structural information but in a spatially-resolved manner. The local ordering may be compositional or structural and  will have some correlation length. A proper determination of these factors is important to understand and control the material behaviour and transformation mechanisms, which is the focus of the current paper.

Structural complexity of materials can be characterized in many ways, such as the atomic variety in alloys, or by the way atoms are correlated and the length a correlation persists.
Since properties of a material are fundamentally linked to its structure, in a heterogeneous material systems, for example, it is important to understand the different parts and relations between different structures.
In sufficiently ordered samples, high resolution transmission electron microscopy (TEM) measurements may be carried out with enough resolution to see individual defects such as dislocation cores and atomically sharp grain boundaries~\cite{williams_transmission_2009}. Although highly informative, these measurements require that columns of atoms persist across the sample, which makes it challenging to gain structural information from heterogeneous disordered regions.
In the most general cases, structural heterogeneity may incorporate a wide range of structural order from distinct, well ordered, crystalline grains all the way to amorphous inclusions. In the latter case, and for small nanocrystalline clusters such as crystalline domains embedded in the amorphous material, direct imaging of atomic columns is not possible.

The structure of amorphous materials is statistical, with many different atomic arrangements. The structure of these materials is often described in terms of local structural correlations, for example, a preference for an approximate FCC-like close packing of nearest neighbours around a central atom. The correlations die away as you move to higher neighbour shells because the packing is not perfect and get progressively random at higher distance from a central atom.  The more perfect the packing the further out the correlations persist. The correlation length is a measure of this persistence length and therefore a measure of the degree of order. To get a complete picture of structurally heterogeneous materials it is important to study the nature of such local structural correlations and how they vary as a function of position in the sample.

\begin{figure}
 \centering
 		\includegraphics[width=0.9\linewidth]{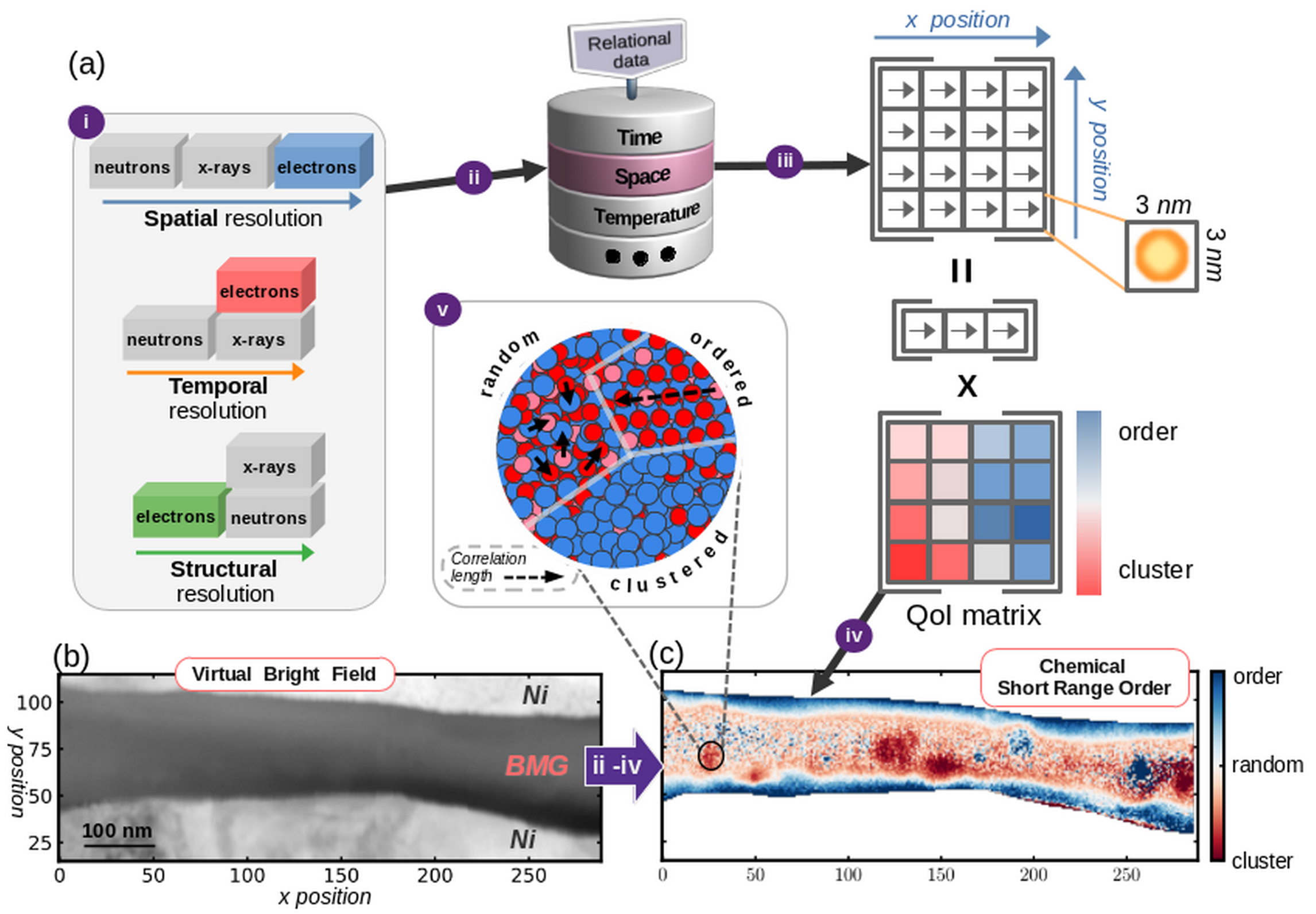}
 \captionof{figure}{(a) An illustration of our approach for extracting structural information from complex systems, which guide the core principles of SNEM. The target is to spatially map the evolution of order in disordered systems as shown in the middle panel. (b) Virtual bright field (VBF) of the heterogeneously disordered sample under study: a single Ni/BMG/Ni laminate from a six-cycle composite accumulative hot-roll-bonded of \thebmg ~ BMG and Ni.  A typical zoomed out perspective of the sample can be found in Shahrezaei \etal~\cite{shahrezaei_synthesis_2019}. As previously shown ~\cite{shahrezaei_synthesis_2019, yavari_situ_1999, fan_ductility_2000, louzguine-luzgin_unusual_2010}, the fabrication technique is expected to provoke a high degree of heterogeneity within the BMG and across the BMG-Ni interface, hence relate to the atomic order illustrated in panel (a-v) and acts as a test system for the SNEM approach. The transformation from panel (b) to (c) shows an example for the implementation of the SNEM methodology, where we show augmentation of order information on-top of a typical virtual bright field image. Panel (c) is a chemical order mapping within BMG layer. The target of the workflow is to be able to identify structural order variations in heterogeneous complex systems, as illustrated in panel (a-v),  such as: a random distribution, an A--B ordering or clustering (blue and red atoms represent different elements of different size and the black arrows shows the correlation-length of the order). The order-mapping workflow, describes: (i) the choice of a probe for a total scattering experiment guided by a resolution criteria (see also \figSS{resolutions}); (ii) collection of relational datasets of diffraction data, where here we focus on spatially-related data; (iii) describes a process for extracting physically-meaningful information using QoIs from the highly-relational data, for example, by the use of a classification algorithms (e.g. a non-negative-matrix factorization, NMF) that use 1D vectors ($\rightarrow$) of ePDFs to reconstruct principle-component vectors and weight matrix for each component; (iv) a QoI matrix that contain order-parameters (e.g., an NMF weight matrix) is then projected into a space (or time) grid to represent order information. In (v) we  represent chemical short range order information.}
 \label{fig:intro}
\end{figure}
%

In alloys, chemical correlations, meaning the a local tendency for alike or different atoms to cluster together or avoid each other, can exist both in amorphous and fully crystalline materials\cite{george_high-entropy_2019}. As schematically shown in panel (v) of \fig{intro}(a), chemical clustering or ordering tendencies, or in short `chemical ordering', can exist regardless to the degree of crystallinity. Therefore, to fully describe a heterogeneous sample, one would need to describe variations in both structural and chemical correlations.

To map order in a heterogeneous or time-evolving sample, the measuring tool need to be with a sufficiently high resolution. For example, when one is interested to spatially-map order gradients, or track processes that include amorphization, nucleation, growth, and chemical ordering, high spatial or temporal resolution would be a requirement. With this in mind, our work claims to disentangle structural and chemical order from a heterogeneous complex system in a nano-meter length scale resolution.

For this purpose, our test-subject is an accumulative hot-rolled composite between a hosting crystalline Ni and a \thebmg bulk metallic glass (BMG) that results in a few 100's of \nm thick BMG laminated in a poly-crystalline Ni matrix~\cite{shahrezaei_synthesis_2019}. More specifically we used a single Ni/BMG/Ni laminate that was thinned to $70\pm15 $ \nm using a focused-ion-beam (\fig{intro}(b)).
\rev{2.11}{The combination of local strain at the interface between the crystalline Ni and the BMG, and the elevated temperature during processing may promote potential diffusion-driven local compositional variations and nano-crystallization~\cite{shahrezaei_synthesis_2019}.} Our aim was to disentangle the underlying structural heterogeneity in the laminated BMG and along the BMG/Ni bonding interface, and trace the local order evolution within the BMG.

To disentangle the structural heterogeneity in the Ni/BMG/Ni laminated test sample, we use electron diffraction. In diffraction, Bragg peaks yield the periodically averaged crystal structure. Deviations from that appear in the diffuse scattering signal. Total scattering and pair distribution function (PDF) analysis take into account both Bragg and diffuse information, yielding in a histogram of all the atom-atom distances in the material that is a convoluted structural length-scale dependent picture of chemical and structural correlations~\cite{egami2012underneath}.
\rev{2.8}{Structural analysis of similar Zr--X based BMG systems was done via x-ray diffraction~\cite{stiehler_effect_2022, wu_multiscale_2017, tong_structural_2016}. However, lacking the spatial resolution electron-diffraction can offer, structural evolution at the nano-scale is impossible.}


The strong electron beam-matter interaction makes electrons the ideal probe when structural information from small volumes or dilute mass (e.g. vapour phase or isolated nano-particles) is desired.
And indeed, total scattering and electron PDF (ePDF) studies of vapour and ultra-thin amorphous layers date back to the 1940's and 1950's. However, for the same reason, ED became challenging for fully quantitative total scattering and PDF analysis on condensed matter due to more significant multiple scattering effects~\cite{egami2012underneath, anstis_investigation_1988}.  
\rev{1.3}{Nevertheless, previous studies showed~\cite{anstis_investigation_1988, ankele_quantitative_2005, mu_mapping_2019} that amorphous materials do not suffer from multiple scattering as much as crystalline ones, meaning that effects on peak position, ratio and shape are benign, even for samples as thick as 150 nm. 
For crystalline samples, besides controlling the fabrication thickness, automated precession-averaging~\cite{vincent_double_1994, portillo_precession_2010} further corrected multiple scattering effects to gain reliable quasi-kinematical ED.}
The introduction of improved automation in ED data collection allowed data to be acquired as a function of position (2D real-space) and scattering momentum (2D reciprocal-space), which introduced a \nm spatially-resolved TEM-based diffraction method known as 4D-STEM~\cite{ophus_four-dimensional_2019}. 4D-STEM has been used to obtain reliable structural information at the \nm scale. 4D-STEM-like measurements often focus on well ordered systems, with clearly-present Bragg peaks~\cite{ophus_four-dimensional_2019, laughlin_physical_2014} where matching diffraction patterns to a known structure to generate, for example, orientation, strain field or geometrically-necessary-dislocation maps~\cite{ozdol_strain_2015, pekin_optimizing_2017, hansen_investigation_2020, kashiwar_situ_2021}.

A pioneering set of works~\cite{mu_radial_2016, mu_mapping_2019, liu_tracing_2020, mu_unveiling_2021, savitzky_py4dstem_2021} showed that when adding PDF to the analysis pipeline of 4D-STEM experiments, one can map amorphous and nanocrystalline phases, beyond the elemental mapping regularly done, for example, using EDX. In these works the ePDFs were used as structural ``fingerprints" to do the mapping. Since it has been previously shown that ePDFs can be quantitatively analysed~\cite{abeykoon_quantitative_2012, gorelik_total-scattering_2015, hoque_structural_2019}, it should be possible to treat the 4D-STEM-generated ePDF data in as similar quantitative manner to map continuously varying structural features in a structurally heterogeneous sample.

Here we present a general concept for extending the current state of the art in structural imaging using transmission electron microscopy as `SNEM', which stands for Scanning Nano-structure Electron Microscopy. A successful SNEM is one that is able to capture a wide range of structural and/or compositional correlations, together with a wide range of correlation lengths (from amorphous to nano-crystals to bulk single crystal), in a spatially resolved manner with \nm spatial resolution. \rev{1.4}{SNEM's purpose is not to resolve from electron diffraction pattern complete structural models, but to follow the evolution of structural motifs in complex systems.}
More specifically, we characterize SNEM as any structurally-resolved scanning transmission electron microscopy method that can result in imaging of structural features at the \nm structural and positional resolutions, and image them as they evolve across a heterogeneous sample in a continuous manner.

To represent continuity in SNEM and draw maps of structural features from diffraction patterns we use a simple but extremely powerful approach: the definition, computation and mapping of Quantities of Interest (QoIs).
QoI refers to any scalar quantity that can be computed directly from the signal, and represents a valuable physical property about the sample. QoI's are widely used for generating maps in various microscopies. For example, the VBF map in \fig{intro}(b) is actually such an example, where the QoI is proportional to the number of counted photons coming from the fluorescence screen due to transmitted electrons that hit close to the beam center.

Since QoI's are mathematically-derived quantities and depend on the analysis pipeline that is used, the number of such quantities that can be mapped is large.  It can even include quantities derived from modelling; for example, fitting models to ePDF data and mapping refined model parameters~\cite{jacques_pair_2013}. SNEM maps strive to contain selected scalar quantities that represent actual structural features, which makes structurally- and chemically-significant QoI harvesting the heart of SNEM. Here we use both 2D diffraction patterns and the reduced 1D-ePDFs as our signal from which QoI's are extracted to generate 2D SNEM maps and discover underlying structural features as demonstrated in \fig{overview}.
\label{sec:intro}
\section{Results and Discussion}\label{sec:results}
\subsection{Overview of the SNEM workflow}
\begin{figure}
 \centering
 		\includegraphics[width=0.8\linewidth]{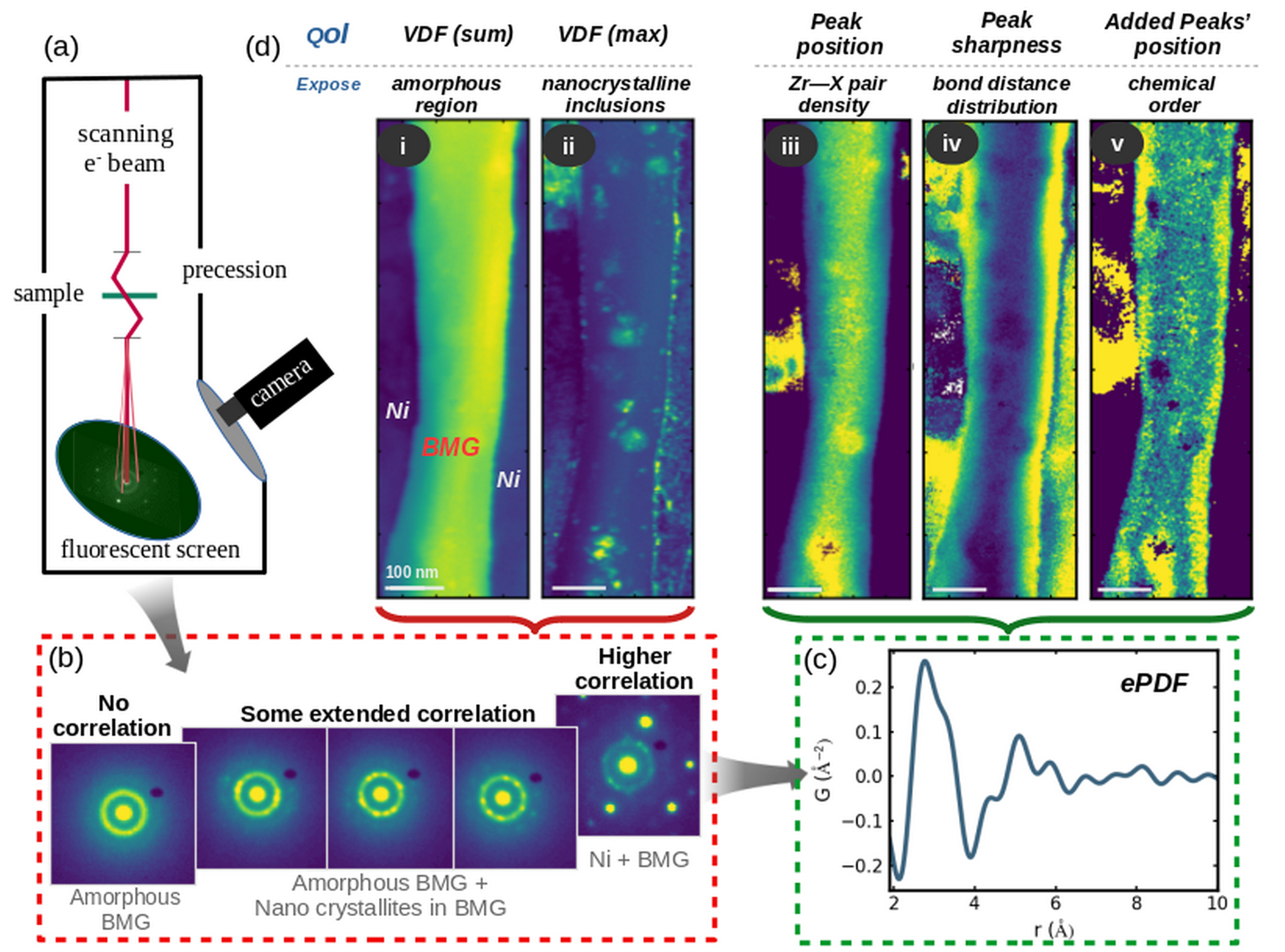}
 \captionof{figure}{An overview summary for SNEM imaging a Ni-BMG-Ni laminate. (a) General schematics of the precession-assisted 4D-STEM data collection setup with a fluorescent screen and a front camera as a detector. (b) Typical distortion-corrected DP's taken from different parts of the sample where one can identify ring patterns from the amorphous BMG, patterns with  extended correlation from nano crystalline inclusions (high intensity spots in rings), and patterns with long-range correlation as shown by apparent intense Bragg peaks coming from regions closer to the Ni-BMG interface. (c) An ePDF that was reduced from a single diffraction patten collected from the the BMG region. (d) SNEM maps using QoI's that are harvested from the DP's or the reduced ePDFs, where a higher/lower QoI value is colored scaled to be more yellow/dark-blue color: map (i) and map (ii) are virtual dark field (VDF) images of the total intensity and the maximum intensity found in an annulus that surrounds the amorphous ring, respectively. Maps (iii) and (iv) use the peak position and sharpness around $r \sim 2.7$~\AA~ as a QoI. Map (v) uses the sum of the first two peaks' position within the r-range 2-4~\AA.
The SNEM maps in (d) demonstrate structural information, including: (i) resolve crystalline from amorphous regions; (ii) detect precipitants with higher-structural correlation (i.e., nano crystallites) in the amorphous BMG; (iii) highlight regions with different pair-correlation density; (iv) map out regions with different distribution of bond-lengths; (v) distinguish between regions with a different chemical short range order.
The maps in panels (d)(ii-v) are repeated and further discussed in \fig{VDF}, \fig{gauss}(b) and \fig{structural}, respectively. Details about the generation of the QoI maps is found in \sect{get_qoi} and along the Results section.}
 \label{fig:overview}
\end{figure}

We start by presenting an overview of the SNEM workflow, which is summarized in \fig{overview}. There follows a more extensive discussion of the resulting SNEM maps (\fig{overview}(d)), and the principles that guide us toward making maps to extract the most scientific insights.  Technical experimental details for  generating the SNEM images are found in the Methods section in the Supporting Information.

\fig{overview}(a) summarizes the data acquisition process. The sample is a bulk accumulative hot-roll bonded
\hfill \break [\thebmg]-[Ni] (BMG-Ni) composite, from which a $\sim 70 \pm 15$~nm thick FIB lift-out of a single Ni/BMG/Ni laminate was used for imaging (\sect{sample_preparation}). A SPED measurement (\sect{data_collection}) was carried out using a front camera that photographed a fluorescence screen detector (\sect{diffraction_pattern_generation}). A beam spot-size of 2~nm was used with a step-size of 3~nm. \rev{2.2}{Electron path Monte Carlo simulation (\figSS{casino}) showed that a 3~nm step size will fall within the lateral resolution of the beam at the exit from the sample.} $289\times 132$ diffraction patterns were collected as our raw data. Due to the data-acquisition geometry, the diffraction patterns were geometrically-distorted, which had to be corrected. We used the commercial `ASTAR' software for the course geometric correction of the diffraction patterns (\sect{distortion_correction}).

Typical distortion-corrected diffraction patterns (DPs) from the BMG layer are shown in \fig{overview}(b). The DPs show both large and subtle differences and encode spatially resolved structural information we will harvest in the form of QoI's.
The DPs show purely-amorphous signals (broad rings), an amorphous BMG mixed with nano-crystalline inclusions (ring with higher-intensity spots in it), a mixture  of crystalline-Ni and amorphous-BMG diffraction close to the BMG/Ni interface, and regions of single or multi-crystal nickel (not shown).
The DPs are used as inputs  to data analysis pipelines that are built in a home-written software `MiniPipes': a configurable Python-based data-analysis pipeline framework (\sect{software}). MiniPipes allows configuring an analysis pipeline for a single DP via a graphical user interface, and applying the configured pipeline on the rest of the DP's resulting in the SNEM maps.

Since our focus is to obtain structural features coming from the BMG layer, we focus our analysis configurations on reducing and extracting QoI's from the diffuse-scattering rings in each pattern. Therefore, QoI's generated from regions that show only (or majorly) Ni-diffraction spots, such as those shown in \figSS{calib} and \figSS{automask}, are neglected for the sake of the scientific discussion. Nevertheless, the crystalline Ni peaks are used for the coarse distortion-correction as well as calibration of the momentum-transfer ($Q$) for each DP. More details are found in \sect{distortion_correction} and \sect{calibration}.

After distortion correction and calibration, the analysis pipeline consisted of these steps: (i) masking, (ii) azimuthal integration to gain a 1D-DP, (iii) reduction and Fourier-transformation of the 1D-DP to obtain the ePDF and (iv) extraction of a QoI from the ePDF.
We note that QoI's may be defined at each step of the analysis, including from the raw data in the form of, for example, virtual dark field images.  The conceptual details of steps \textit{i-iv} are presented below, where the full details are discussed in the Methods section in the Supporting Information.

\rev{1.2}{When we used ePDF to extract a structural QoI's, we applied an auto-masking algorithm on every distortion-corrected DP before azimuthally-integrating the DP images. This step was important in particular to isolate intensities that originate from the amorphous part of the sample. As shown in \fig{overview}(b), the amorphous signal from the BMG is occasionally mixed with Bragg peak intensities from nano crystalline inclusions within the BMG, or Ni peaks close to the Ni/BMG interface. The algorithm for masking static pixels and occasionally-appearing defective pixels or unwanted Bragg diffraction intensities is further explained in the \sect{get_automasked} and demonstrated in \figSS{automask}. Following the masking step we azimuthally integrate the masked DP images.}
To get an accurate $Q$ mapping for the azimuthally-integration (\sect{get_iq}), a calibration process (\sect{calibration}) was used. We note that since calibration can be sometimes challenging, and since in microscopy one may be interested in relative changes without insisting on absolute values, limited but valid qualitative scientific insights can be also drawn with a less accurate or even no calibration at all.  The 1D integrated DPs were then reduced and Fourier-transformed to get a PDF, $G(r)$  (\sect{get_pdf}), as shown in \fig{overview}(c).

Multiple physically-significant QoIs can be defined and computed (\sect{get_qoi}). Indexing each scalar QoI with the spatial position, can then generate SNEM images as shown in \fig{overview}(d). As such, the SNEM images allow us to visualize different aspects of the structural heterogeneity. The maps in \figs{overview}(d)(i-ii) (red box) use extracted QoI's from DP's, while \figs{overview}(d)(iii-v) (green box) use extracted QoI's from ePDFs. Each SNEM image comes from a unique contrast signal that emphasizes different structural features and exposes different aspects of the structural heterogeneity.

Map \fig{overview}(d)(i) uses as a QoI the total intensity in an annulus surrounding the main diffraction ring (see \figSS{qois}(a)). Map \fig{overview}(d)(ii) uses the same annulus mask but extracts the maximum intensity in the annulus. \fig{overview}(d)(i) clearly distinguishes between the crystalline Ni and the amorphous BMG, \fig{overview}(d)(ii) exposes regions with nano-crystalline inclusions within the BMG, which result in stronger diffraction peaks due to more ordered clusters present within the probed area, i.e., nano-crystalline inclusions.

Simple, but extremely valuable QoI maps are for a PDF peak position and sharpness extracted directly from the PDF data, as shown in \figs{overview}(d)(iii) and (iv).  These contain information about the average interatomic neighbour distances and bond distribution in the first coordination shell (CS).
Since in our system the atoms have different atomic sizes, with Zr being the largest, map \fig{overview}(d)(iii) yields information about the Zr--X pair density with respect to X--X pair density (X is any other atom than Zr). Since bond-density distributions generally correlate with the local chemical composition bright-yellow regions roughly reflect a more Zr-rich area. Later we show that this conclusion is consistent with an EELS measurement that yields a spatially-resolved elemental analysis from the same region on the sample.

We further note that the SNEM measurement gives not only information about the average composition vs. position, but also any variations in local chemical and structural order even when the composition is not varying.   Such effects are are often met in multi-component systems and introduce heterogeneity in the local order~\cite{george_high-entropy_2019}.

In this regard, the distribution of bond-lengths in the first CS, as shown in map \fig{overview}(d)(iv), shows regions with sharper (more yellow) or less sharp (more dark-blue) peaks that correspond to the distributions of bond-lengths. We see, for example, that closer to the Ni-BMG interface the bond-length distribution is narrower than in most regions within the BMG core. This can indicate a reduction in the variety of atomic species that contribute to the bond-length distribution, such as when we have a Ni-rich region closer to the Ni-BMG interface. In the middle of the BMG, the wider bond distribution regions appear as dark-blue regions.

A slightly more advanced QoI map that uses both peaks' position within the range $2<r<4$~\mAA is shown in \fig{overview}(d)(v) and yields information about the chemical short-range-order (SRO), meaning the tendency of the different elements in the structure to cluster (form X--X and Y--Y bonds) or mix (from X--Y bonds). Moreover, the vast amount of spatially-related diffraction patterns can be used to extract chemical SRO maps using principle component analysis, such as non-negative matrix factorization.

In the following section we deal with the physical principles that justify our choices of QoIs for the generation of the different SNEM maps to come. Our leading principle is scientific insights one can draw from them about structural ordering in glasses when gradual and abrupt compositional variations are involved. The implementation of the SNEM maps for understanding better the exemplary BMG/Ni system will then be discussed, such as structural ordering in laminated metallic glasses, possible binding mechanism between amorphous and crystalline interfaces and their link to the local chemical composition.

\subsection{Identifying significant QoI's}
QoI's are simply constants that can be computed from the measured data, and thus varied and limited only by the imagination of the experimenter. One of the goals is to identify QoI's that will represent direct and interpretable local structural quantities that vary across the sample. In our case, we focus on an amorphous material with spatially varying compositional and structural correlations, including a form of nano-crystalline inclusion~\cite{shahrezaei_synthesis_2019}. Therefore, our goal is finding QoI's that provide information mostly about the short-range structural and chemical order.

Although the choice of QoI's is a matter of an exploratory process, since SNEM is expected to be evaluated against the scientific insight it potentially holds, we need to understand the conceptual significance of structural order in BMGs and BMG-crystallite composites. To make this connection, we start with a short overview on the challenges and unknowns in BMGs.

Although most metals tend to crystallize even with high cooling rates, BMGs can freeze in the amorphous structural phase at relatively low cooling rates ($<1000$~K/s), allowing amorphous alloys with bulk thicknesses ($> 1$~mm) to form~\cite{jafary-zadeh_critical_2018}. BMGs  have superior strength, hardness and wear resistance than traditional crystalline alloys~\cite{kruzic_bulk_2016}. However, the use of BMGs in engineering applications is limited due to spontaneous localized strain softening behavior that can lead to unpredictable catastrophic failure. Strain softening is usually blamed on shear-bands that create shearing highways, which after nucleation, propagate during applied loads and progressively weaken the alloy's resilience~\cite{greer_shear_2013}. The atomistic origin for shear-band nucleation and propagation is an area of active study~\cite{wu_atomistic_2021,yang_susceptibility_2019, sopu_atomic-level_2017, takeuchi_atomistic_2011, ye_atomistic_2010}, and involves attempts to understand the importance of chemical short range order (CSRO) and alternative local structural relaxation pathways. Similar to crack-propagation, there is a critical cross-section under which shear-bands are suppressed and deformation mechanisms are accommodated by collective atomic rearrangements known as shear transformation zones (STZ), in which homogeneous deformation can take place~\cite{volkert_effect_2008, jang_transition_2010}. Computational and experimental work has shown that homogenous deformation is possible in amorphous/crystalline nanolaminates, where homogenous deformation is accommodated by a collaborative interaction of STZs and dislocations~\cite{kim_nanolaminates_2011, brandl_structure_2013, cui_hardness_2015, gupta_dislocation_2017, cheng_design_2018, shahrezaei_synthesis_2019}.

The BMG in our sample is embedded in a crystalline matrix and has undergone significant deformation as part of the hot rolling. Of particular interest is what happens at the interface between the BMG and the matrix after this process. For example, does it  retain an atomically sharp interface? Furthermore, has the processing resulted in heterogeneities such as chemical segregation or nucleation of stable crystallites. Since order in glasses is often undesirable, potentially causing shear-band nucleation centers and brittleness~\cite{kruzic_bulk_2016}, one thing we would like to understand is whether there is a connection between specific local chemical order to the tendency to form emergent structural features. To get at the structural order we can explore the first coordination shell to see the distribution of bonds, such as the tendency for clustering of similar atoms versus a random distribution, information that is accessible in the PDF in the nearest neighbor peak. Then, we would like to link the distribution of nearest neighbors to the correlation length and try to understand if a tendency towards crystallization follows a particular chemical or bond-density distribution.

\subsection{SNEM analysis of the hot-rolled BMG sample}\label{sec:2_3}
\begin{figure}
		\centering
 		\includegraphics[width=0.7\linewidth]{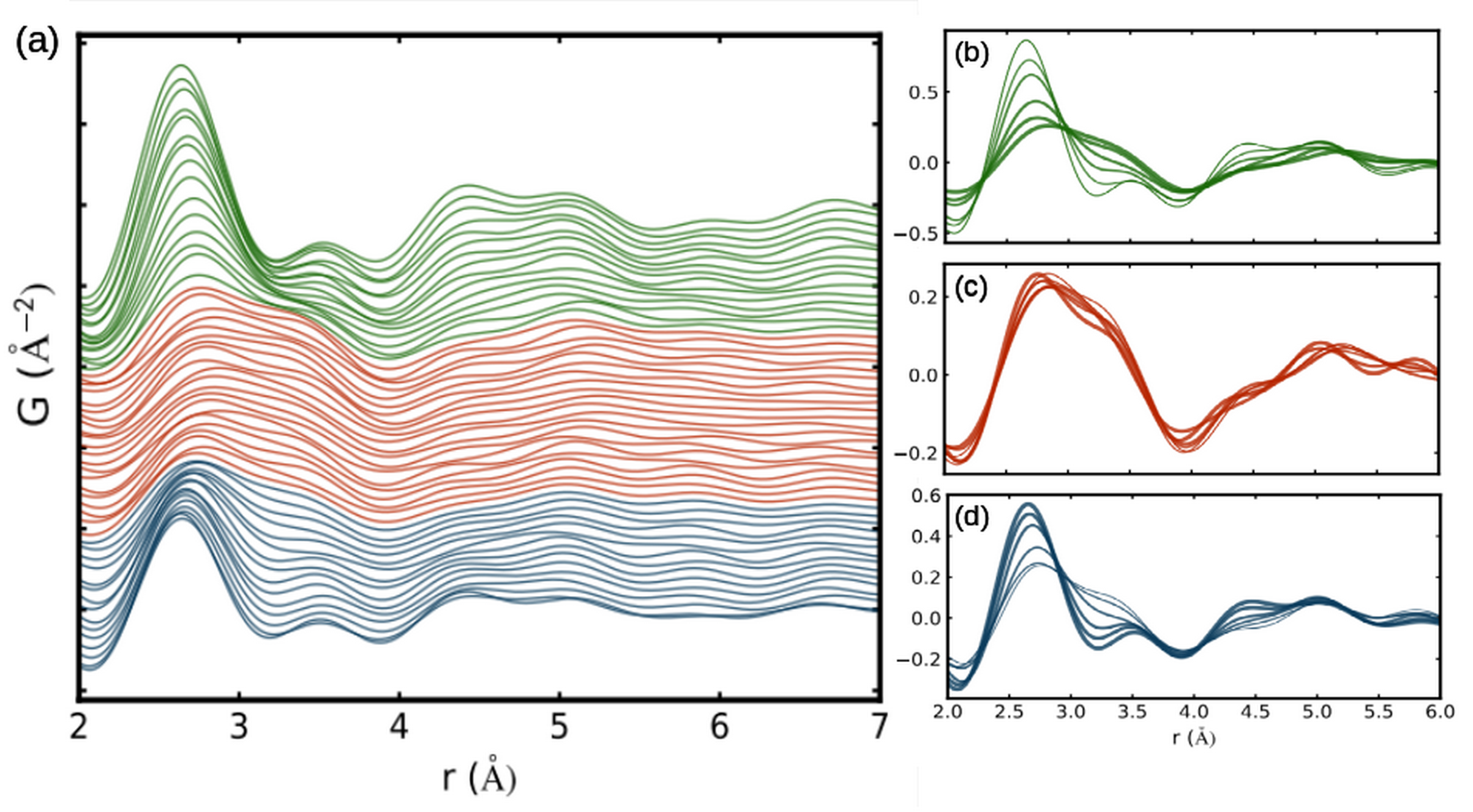}
 \captionof{figure}{(a) ePDFs from points across the BMG layer. The ePDFs were taken from a cut where no nano-crystalline inclusions were observed within the BMG (cut $x=75$) and within the amorphous BMG layer ($55 < y < =100$). The different colors represent regions of the BMG where the PDFs are qualitatively different. (b)-(d) Representative PDFs (every third plot within the $y$ range) from the differently colored regions plotted without an offset for direct comparison. In (b)-(d) the line thickness varies from the thickest at the lower $y$ index to the thinnest at the highest $y$ index within each stack.}
 \label{fig:waterfall}
\end{figure}

We begin with an overview of the general trend of the ePDFs across the BMG layer. Choosing a cut where no intense peaks due to nano-crystalline inclusions are present to represent a purely amorphous set of ePDFs (see \fig{overview}(b)), we plot a waterfall of ePDFs, shown in \fig{waterfall}(a) and corresponding stacked ePDFs from each one of the three characteristic regions across the BMG, shown in \figs{waterfall}(b)-(d). We note that although the characteristics of the three regions is general for other cuts across the Ni-BMG-Ni region, other cuts may have different profiles. Before getting into the specifics of each ePDF, we first address the most pronounced differences between the center (red) and edge (blue and green) regions.

%

The red region in \fig{waterfall} is the center of the BMG.  The PDFs are similar in this region, suggesting that in this particular cut the local structure is not evolving across the middle part of the BMG block.  The blue and green regions show a smooth transition from the PDF of the center-region to a distinct PDF at the Ni-BMG interface. Within the first nearest neighbor region of the PDF (an $r$-range of 2-4~\AA) as we move from the middle to the edge of the BMG, intensity is lost from the $r=3.25$~\AA\ region, whilst a sharp peak emerges on the low-$r$ side of this region and shifts to lower-$r$, finishing at $r=2.65$~\AA~ at both the bottom and top edges of the BMG.
Other features in the PDF can be seen to be changing systematically with position. For example, at higher $r$ distances ($4< r < 6$~\AA), i.e., higher order CSs, we find in the edge regions (blue and green) an evolving peak around $r=4.5$~\AA. We also find that the peak around $r=5$~\AA\ in the red region varies; however, the changes seem to be abrupt. In principle, our first attempt for extracting structurally-related QoI's can focus on the three above mentioned peaks: around $r \sim 2.7$~\AA, 3.4~\AA\ and 4.5~\AA. For the sake of simplicity, we continue with the first two peaks between 2-4~\AA.

In general in a BMG such as \thebmg\ there will be Ni-Ni, Ni-Zr, Zr-Zr and other interatomic contacts present which will vary in number and length depending on the local composition and the efficiency of the packing. The relevant atomic radii are~\cite{knovel_firm_chemistry_1997}: $r_{Zr}=1.60$~\AA,   $r_{Cu} =1.28$~\AA,   $r_{Ni} =1.24$~\AA,  and   $r_{Al} = 1.43$~\AA, which would yield Ni-Ni contacts as short as 2.48~\AA\ and Zr-Zr contacts as large as 3.2~\AA\ with others in between.  Because of the relatively large amount of Zr in \thebmg\  composition, we expect significant intensity in the PDF at $r\sim 3.2$~\AA\ at the center region of the BMG, and indeed that is the case.  In fact, all the peaks seem to be pushed to slightly higher $r$-values than would be expected in a close packed material, due to the packing inefficiencies inherent in such a mixed component glass~\cite{zhang_origin_2015}.

\begin{figure}
	\centering
 	\includegraphics[width=0.90\linewidth]{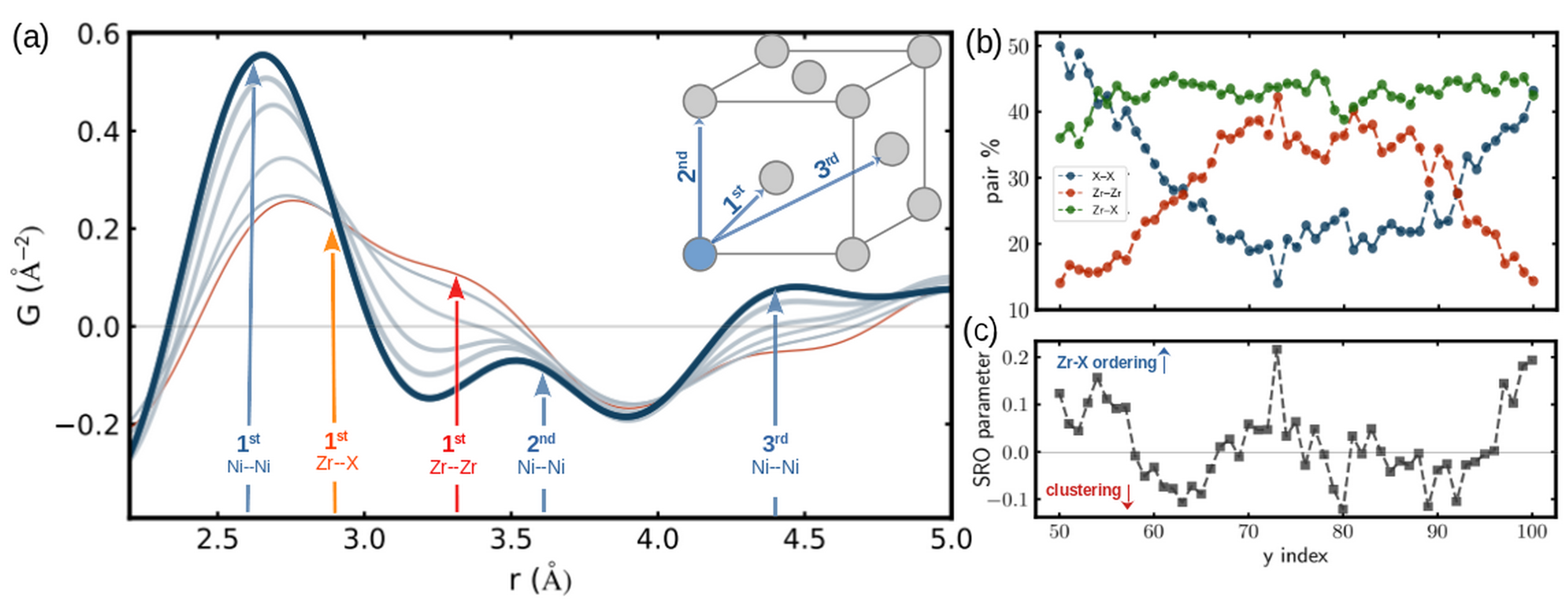}
 \captionof{figure}{(a) ePDFs (replotted from \fig{waterfall}(d)) with an emphasis for the structural evolution between the Ni/BMG edge (thickest blue) and the center of the BMG (thinnest red), where the thicker the line the closer it is to the Ni-BMG interface. The inset shows an illustration of a close packed FCC structure with the indices of the nearest neighbors in the structure. The blue arrows point to the position of corresponding Ni--Ni expected distances if the material were close packed, with a shift of 0.1~\AA\ from the nominal closed-packed Ni (here demonstrated by an FCC structure), as one may expect in an imperfect packing of a glass. The orange and the red vertical arrows correspond to Zr--X (X being mostly Ni or Cu) and Zr--Zr nearest neighbor distances with a similar imperfect packing shift of 0.1\AA. (b) The relative percentage of X--X, Zr--Zr and Zr--X pairs as derived from the refined scaling factors for each pair. The scaling factors are refined during the fitting of ePDFs with the structural model along the same cut presented in \fig{waterfall}(a) (see text and fitting examples in \figSS{pdffit}). Each point corresponds to the selected $y$ indices across the x-cut. (c) The corresponding SRO parameter, $alpha$, using the values in (b) following \eq{sro} and \eq{p_max}.}
\label{fig:coordshell}
\end{figure}

It is interesting that, as the edge of the BMG is approached, intensity disappears in the $r=3.2$~\AA\ region and starts to peak up at around $r\sim 2.7$~\AA, which is close to, but slightly higher-$r$ than the ideal close-packed Ni-Ni distance, $r_{nn}$.  In close-packed nickel, the second neighbor distance is at $\sqrt{2} \cdot r_{nn}$ and the third neighbor at $\sqrt{3} \cdot r_{nn}$.  This is illustrated in \fig{coordshell}(a).

%

If the edge region is very rich in Ni, allowing it to become more close-packed (though still amorphous globally) we might expect to see intensity in the PDF appearing at around $r=2.5$, 3.53 and 4.33~\AA, the positions of Ni in its fcc structure, as we move towards the edge of the BMG.  Such a trend is observed, albeit with the observed $r_{nn}$ expanded by $\sim 0.1$~\AA\ or so, indicated by the arrows in \fig{coordshell}(a).\footnote{The small shift in the Ni-Ni nearest neighbor distance is expected because of the alloying; however, we add a note of caution that imperfect calibration of the effective sample-detector distance may also cause a shift in peak position. We also note that a sharp PDF peak evident at $r=3.5$~\AA\ has a contribution from termination effects due to the sharpening nearest neighbor Ni peaks and the limited $Q$-range of our measurement.  It undoubtedly reflects intensity in the PDF coming from Ni-Ni second nearest neighbor peak, but the termination ripple may cause a shift in the position of the peak maximum from where it should be and we caution against over-interpretation of this feature.} Overall, we see that the amorphous material is becoming very nickel-rich and somewhat close-packed in this transition region at the boundary with the crystalline nickel.

As we go deeper into the BMG, Zr--Zr and Zr--X pairs (X is Ni, Cu or Al) become more prominent in the ePDF profile as evident by the grey and red curves in \fig{coordshell}(a), in which an increase in intensity is evident around $\sim 3.2$~\AA\ (Zr--Zr) and $\sim 2.8$~\AA\ (Zr--X). To understand the distribution of pairs in more detail, quantitative modelling is required. The reason is that the relative intensity of the peaks depends on the atomic number as well as the multiplicity of that type of pair. The corresponding atomic numbers are $Z_{Zr}=40$,    $Z_{Cu} =29$,    $Z_{Ni} =28$,    $Z_{Al} = 13$.

Given our structural resolution for separating between pair-distances contributions in a PDF ($\sim0.39\AA$ - see \figSS{resolutions}), because of the similar sizes of Cu, Ni and Al but a significantly larger Zr, we approximate the signal as coming from a binary [Zr,X] compound that contains Zr--Zr, Zr--X and X--X paired bonds. As a start we assume the bonding forms a relatively closed-packed, FCC-like, arrangement for all three paired bonds. We then model the experimental ePDFs using a sum of three independent FCC structures with effective lattice parameters of $\sqrt{2} r_{nn}$. One can then extract the partial PDFs for each pair (see \figSS{partial_pdf}), meaning a relative contribution of each ``atomic pair'' in the pseudo-binary alloy to the ePDF. In order to estimate the validity of this model to our experimental ePDFs, we fit a few ePDFs from \fig{waterfall} (from edge to center of the BMG) to a structure that contains Zr--Zr, Zr--X and X--X bonds. In this simulation we fixed all the structural parameters (as shown in \tablSS{fit_params}) and released only the scaling factors for each pair to produce a simulated dataset that was fit with our simplified model.
In \figSS{pdffit} we show that by allowing only the lattice parameters and scaling factors to adjust, a fair degree of agreement between the model and the data is achieved (see simulations of the partial PDFs in \figSS{partial_pdf}). Closer to the BMG-Ni interface the agreement between the model and the data is better than in the core of the BMG (compare $y$ index 80 with 50 in \figSS{pdffit}). However, both cases show a very good agreement around the first nearest neighbors. This suggests that closer to the Ni-BMG interface the FCC model correlates to longer distances, while within the BMG-core the second nearest neighbor agrees less with the closed-packing model, which makes sense due to the greater range of elements in the BMG central region, as later verified with EELS (\fig{eels}).

Having validated this approach, we estimate the partial contribution of each pair (Zr--Zr, Zr--X or X--X) to the measured PDFs, doing this for every 5~pixels across the BMG map (corresponding to the PDFs in \fig{waterfall}). We can clearly see the evolution of the individual pairs across the BMG.
The ratio between the scaling factors from the fitting to our model allows us to give a semi-quantitative estimation of the percentage of each pair, i.e. the ratio between X--X, Zr--Zr, and Zr--X pairs. Extracting and normalizing the scaling factors from each pair to a total of 100\% (\fig{coordshell}(b)) we gain insight to the evolving distribution of pairs across the BMG. \fig{coordshell}(b) clearly shows that closer to the Ni-BMG interface (towards $y$-indices 50~and 100) the X--X pair density increase while Zr--Zr pairs decrease, which is expected (and later verified) due to the increase in Ni concentration. Towards the center of the BMG, and as expected from our Zr-rich BMG compound (recall nominal concentration: \thebmg), the Zr--Zr density is higher than that of X--X. Unlike the X--X and Zr--Zr pairs, fluctuations in Zr--X pair density is more subtle which fluctuates around 42\%.

In most cases, except where the X--X pair density is very high close to the edge of the BMG, the Zr--X pair density is higher than the other pair densities. However, it would be interesting to understand whether the local bonding tends towards clustering or ordering, meaning whether atoms tend to coordinate with like atoms (i.e., X--X and/or Zr--Zr clustering) or with dislike atoms (i.e., Zr--X ordering), which we discuss below.

In a perfectly random binary solid solution the number of Zr--X nearest neighbor pairs will be exactly twice that of the lowest among the Zr--Zr or X--X pairs. Given an estimate of the pair concentrations, the short range order (SRO) parameter, $\alpha$, can be defined as~\cite{porter_phase_2009}:
\begin{equation}
\alpha = SRO_{[Zr-X]} \equiv \frac{P_{[Zr-X]} - P_{random}}{P_{max}-P_{random}} = \frac{P_{[Zr-X]}}{P_{random}} - 1
\label{eq:sro}
\end{equation}
where $P$, $P_{max}$ and $P_{random}$ are the actual, maximum and random-case concentration of Zr--X pairs, respectively. The given pair concentration, $P_{i}$, can be equivalently replaced by the pair fractions, which are given in \fig{coordshell}(b). For a given set of pair fractions, the maximum possible number of Zr--X pairs is the existing fraction of Zr--X pairs plus the fraction of the lowest pair density among Zr--Zr and X--X pairs. With a random distribution, $P_{random}$, being half of the possible maximum number of pairs, $P_{max}$, we can then write for a binary system that
\begin{equation}
P_{max} = 2\cdot P_{random} = P_{[Zr-X]} - 2 \cdot  \min\left(P_{[X-X]},~P_{[Zr-Zr]}\right).
\label{eq:p_max}
\end{equation}

Since we know $P_{[Zr-Zr]}$, $P_{[Zr-X]}$ and $P_{[X-X]}$ from our fits (\fig{coordshell}(b)), we use \eq{sro} and \eq{p_max} to calculate the SRO parameter, $\alpha$, which is plotted in \fig{coordshell}(c). As expected, we find that $\alpha$ oscillates around zero, which is the case of a random solid solution. Any value above zero indicate a tendency for a Zr--X ordering around the first CS, while values below zero indicate a tendency for clustering of similar atoms. The effect is small, but there is arguably a weak ``W" shape across the BMG with a small clustering tendency towards the edges (e.g., $y$ indices 58-66 and 85-94), and ordering towards the middle.
Such an insight to the statistical distribution of structural topology around the first coordination shell with a \nm~spatial resolution is unprecedented, and given the observed variations in $\alpha$, it is clear that a \nm~scale resolution is required for observing and understanding the structural ordering in materials such as these BMGs. The SNEM analysis clearly presents a powerful tool for studying structural heterogeneity at the relevant length-scales.

We have limited our quantitative analysis to assessing line-cuts through the BMG. We now turn to making full 2D QoI maps. Our attempts will focus on model-independent QoIs derived from the raw DP as well as quantitative analyses of the ePDFs. Our first attempts will focus on the structural information that is in a limited $Q$- (in DPs) or $r$- (in ePDFs) range. Such 2D QoI maps were presented in \fig{overview}(d) and are explained in greater detail below.
We then use the full data-set and extract similarity QoIs using the Pearson-correlation metric (\fig{pearson}), and extract the partial weights of principle components using non-negative-matrix (NMF) factorization (\fig{nmf}). These turn out to be very useful for mapping regions of similarity and to map the contributions of the different components to the signal. As will be shown, the NMF components resemble the partial X--X, Zr--X and Zr--Zr PDFs, which will allow us to map first NN pair-concentration maps, or `bond-maps', and as demonstrated in \fig{coordshell}(c) and to create a map of the SRO parameter discussed above. We will end with augmenting EELS-based elemental mapping, and will focus on understanding the complementarities and differences between heterogeneities in compositional and structural space.

\subsection{Exploring inclusions in the SNEM images}
\begin{figure}[t]
	\centering
 		\includegraphics[width=0.50\linewidth]{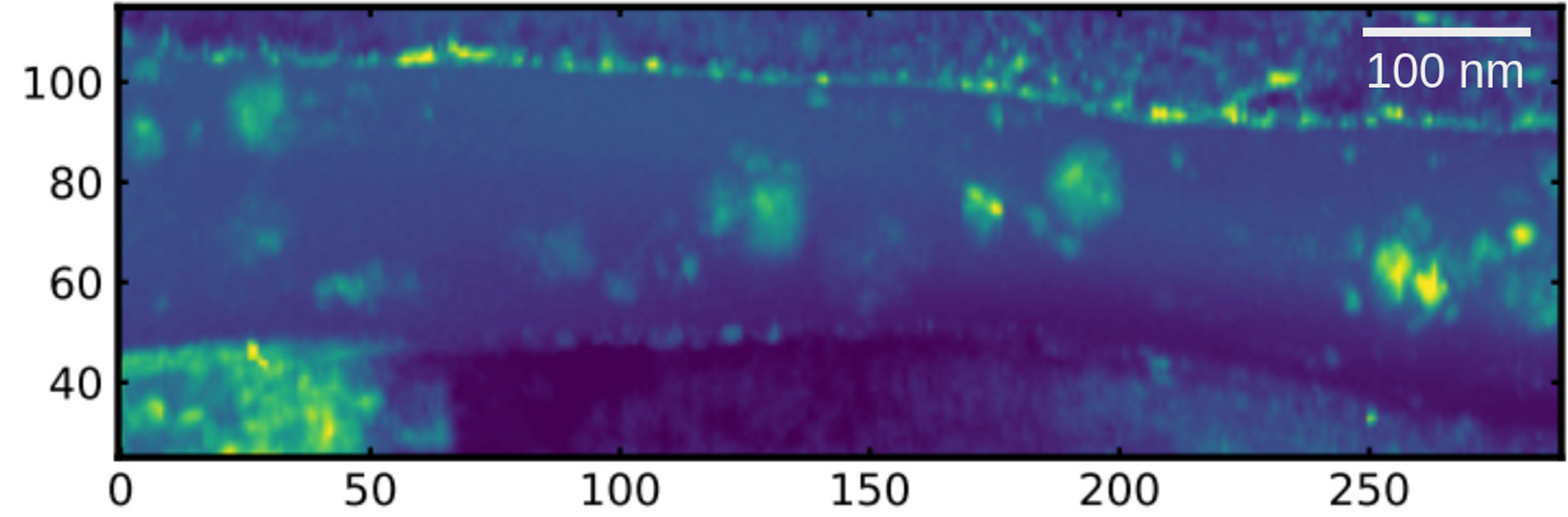}
 \captionof{figure}{Virtual dark field image in which each pixel is the maximum intensity found in an annulus that surrounds the ring (amorphous) pattern (see \fig{qois}(a)). Bright (yellow) regions indicate presence of a nano-crystalline inclusion that creates a more intense diffraction spots in the ring. Similar maps that are sensitive to the crystallite orientation are shown in \figSS{VDF_point}. For this study, however, we will be interested in whether a nano-crystallite is present rather than its orientation. }
 \label{fig:VDF}
\end{figure}

\begin{figure}
\centering
 		\includegraphics[width=0.9\linewidth]{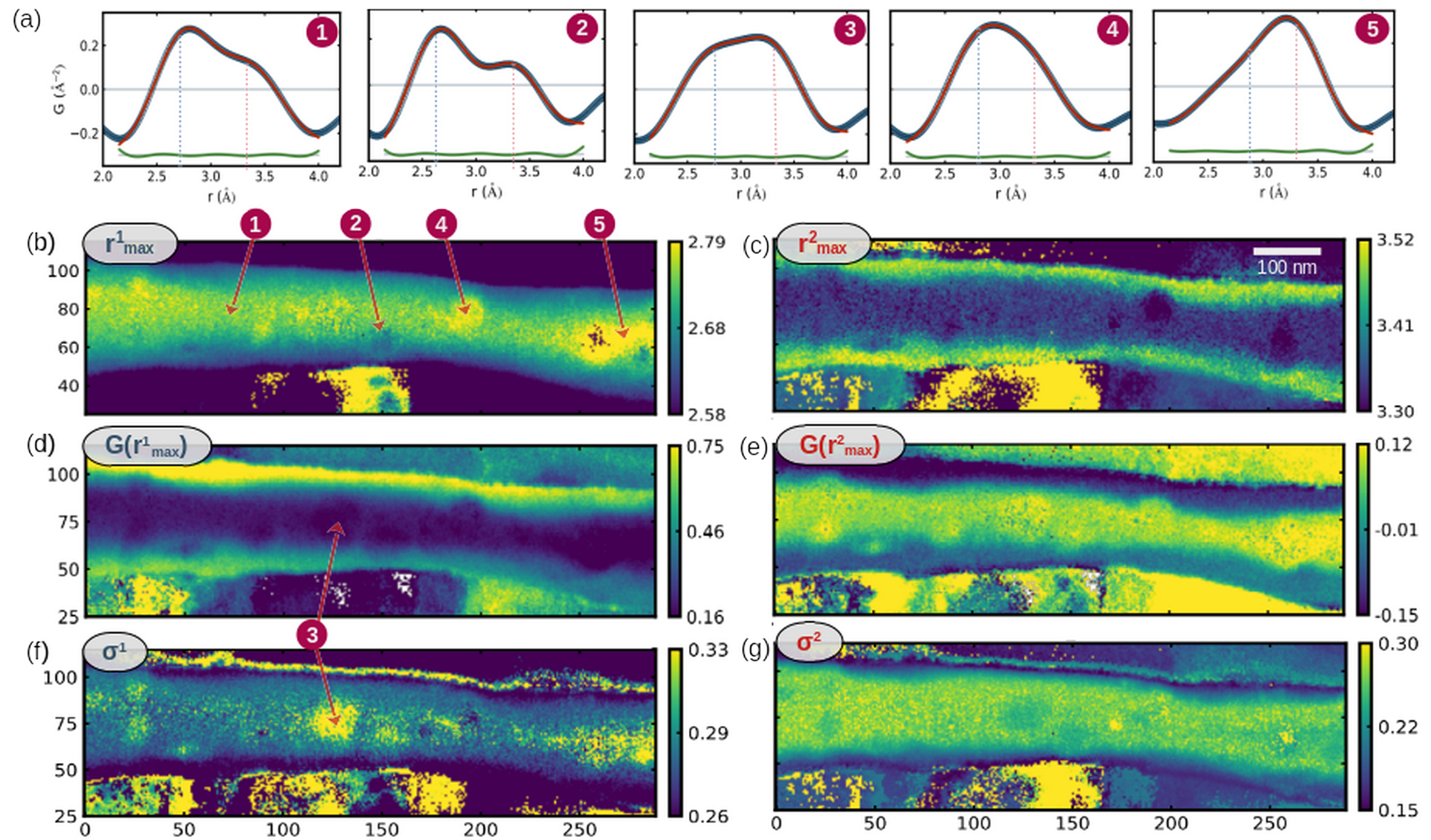}
 \captionof{figure}{(a) Characteristic examples for a double Gaussian fits from regions that show contrasts in the subsequent SNEM images (b-g). The blue and red vertical dashed lines in (a) represent the peak position, as derived from the fitting. The fit is represented as a red curve, and the difference between the fit and the data is plotted in green. (b)-(g) Derived SNEM images following a double-Gaussian fitting, where the selected QoI is mentioned in the gray bubble: (b,c) are the peak positions, i.e., $r_{max}^{i}$; (d,e) are the peak values, i.e., $G({r_{max}^{i}})$; (f,g) are the peak widths , i.e., $\sigma^{i}$.  A global length scale bar is presented in (c).}
 \label{fig:gauss}
\end{figure}

The maps shown in \fig{overview}(d) from these very simple analyses of the PDF (the value and position of the first neighbor PDF peak) already reveal a lot of structural information about the sample and heterogeneities present there.  For example, as well as variations in composition and structure from the edge of the BMG to the center that we have discussed above, circular inclusions are evident at various places, and an elongated bright region at the bottom of the image. It is possible to make a QoI map from the data at any point in the processing, and indeed more conventional virtual dark field images are QoI maps that can be useful for imaging nano-crystallinity in the sample. The particular diffraction pattern QoI we used was the maximum pixel intensity within a circular annulus on the diffraction pattern that coincided with the first diffuse diffraction ring (see \fig{qois}(a)). Because the nanocrystallites give slightly sharper peaks than the amorphous material, intensity from them appears as brighter spots in the diffraction ring.  Our QoI captures this feature and, just like a conventional VDF image, maps their locations.  This is shown in the VDF(max) panel of \fig{overview}(d) regions, and in greater detail in \fig{VDF}. A simple visual comparison between the VDF(max) and PDF peak position and width maps in \fig{overview}(d) (also \fig{VDF} and \fig{gauss} bellow) shows that the circular objects in the PDF maps coincide with the presence of nano-crystalline inclusions.
%

We now explore how we can learn more about the local chemistry and structure in these inclusion regions, and the glass regions around them, by studying the PDFs in those regions.

At this point we have identified that the first two peaks in the PDF (2.1-4.0~\AA) contain valuable information about short-range local structural order, as well as information about the density of Zr--Zr, Zr--X and X--X bonds. These different bond-length distributions cannot be directly resolved because they are intrinsically broad in the glass system, and also our $Q_{max}=8.1$~\AA$^{-1}$ is somewhat low (resulting in a real-space resolution limit of  0.39~\AA\ ($\mathrm{\Delta r=\frac{\pi}{Q_{max}}}$~\cite{farrow_nyquist-shannon_2011}).  However, more information can be extracted by modelling the PDF, such as demonstrated in \fig{coordshell} and \figSS{pdffit}. Rather than fitting the entire dataset and dealing with the discussed-above challenges it comprise, we take the simpler approach of fitting multiple Gaussians to the compound peaks in the region $2.1<r<4.0$~\AA , as shown in \fig{gauss}(a) (for further details, see \sect{get_qoi}).
Spatial SNEM maps can then be made for each refined parameter from these fits (see \tablSS{params}(iii) for details). The resulting SNEM images, are shown in \figs{gauss}(b-g).

First, we note that the shape of the closest shell PDF region varies greatly between  different inclusions, as is evident in \fig{gauss}(a). A PDF from the non-inclusion region is labelled (1) and shown in \fig{gauss}(a)(1).  This is similar to the PDFs shown in \fig{waterfall} (red curves) which are from a cut across the BMG that doesn't go through an inclusion.  \fig{waterfall} also serves to show the degree of reproducibility in the PDFs when the local structure is not changing.  By contrast, all of the PDFs from inclusions are significantly different from each  other, with the relative weight of peak intensity on the left and right side of the unresolved peak multiplet, varying considerably even though all the inclusions we consider are all located in the region close to the central axis of the BMG.  \rev{2.9}{As mentioned earlier, the first coordination shell project information about the chemical order,  and, to some lesser extent, the packing symmetry (packing symmetry will be emphasized more in the second and third coordination shells). Variations in composition, local chemical order and correlation length, will affect the first coordination shell.} This shows unambiguously that the local structure (and chemistry) varies from inclusion to inclusion and between the inclusions and the amorphous matrix.

%

Analysing \fig{gauss}(b), we find that the values of the first-peak position, $r_{max}^{1}$, spans mostly between 2.65-2.75~\AA, which fits the range of a mixture between X--X and Zr--X. Since the contribution to the first peak comes only from X--X and Zr--X, a primary contribution to a peak shift to higher $r$ values (indicated by yellow regions in \fig{gauss}(b)) would be a higher density of Zr--X pairs, where a secondary contribution would be a higher Al concentration. We will see later (\fig{eels}(d)) that the Al concentration stays roughly the same throughout the BMG, which suggests that regions in the $r_{max}^{1}$ map that are yellow have more Zr-X pairs.  This will increase in regions that have higher Zr and decrease in regions with less Zr, assuming random packing.

\begin{figure}
	\centering
 		\includegraphics[width=0.70\linewidth]{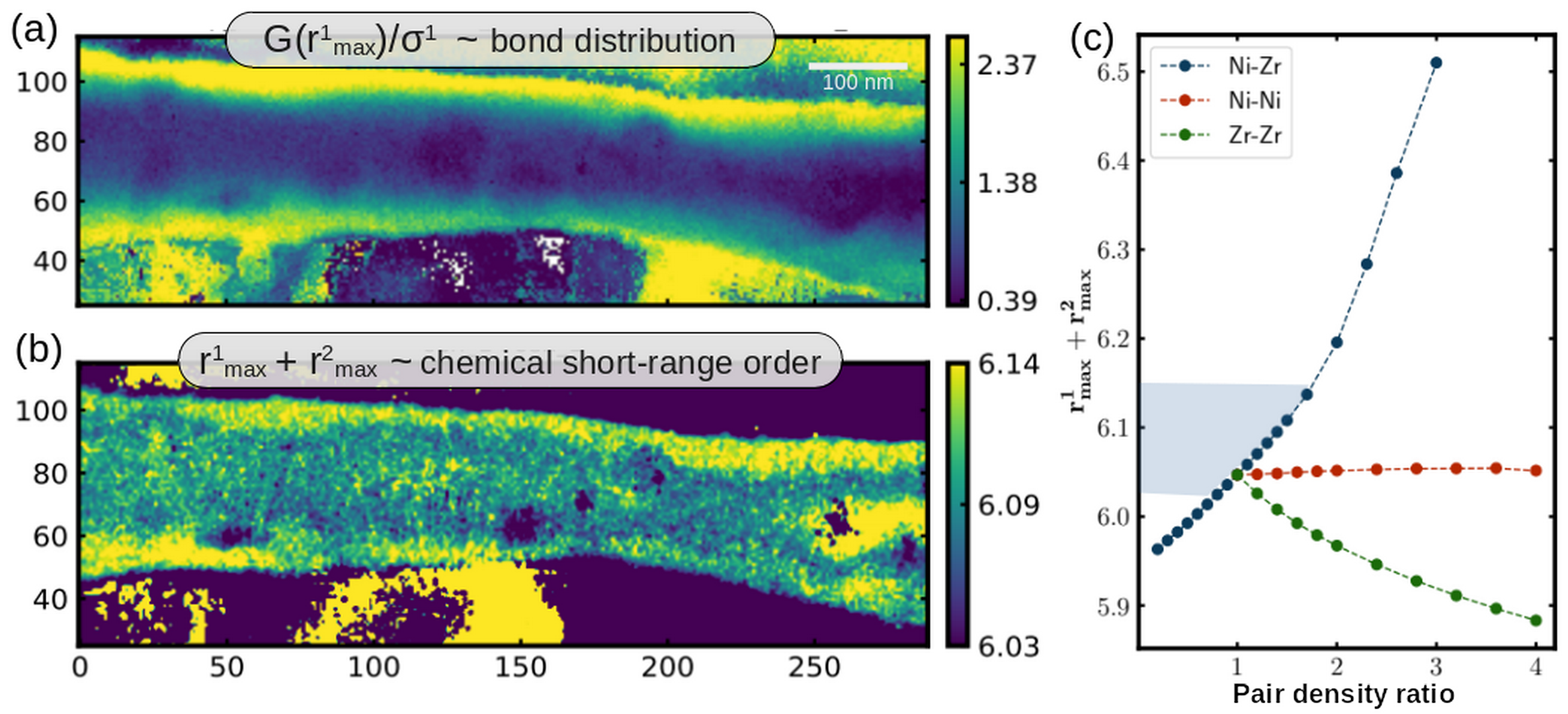}
 \captionof{figure}{An advanced set of QoI maps that expose specific structural aspects. The maps are generated using the fundamental QoIs that were derived for \fig{gauss}. Map (a) is the first peak sharpness and it is intended to expose variations in bond-distribution in the BMG, where darker regions refer to a greater bond distribution. Map (b) is the sum of the first peak positions and it is intended to expose chemical SRO, in particular regions where chemical clustering is likely to be present. According the simulations shown in (c), clustering, that is characterized by an excess of X--X or Zr--Zr with respect to Zr--X, would be characterized by reduction the summed value of the peak centres, $r^{1}_{max}+r^{2}_{max}$. In (c) the resulting $r^{1}_{max}+r^{2}_{max}$ values are from simulations (\figSS{r1r2sim}) using a closed packed set of pairs (the same model used for fittings in \figSS{pdffit} and simulation of the partial PDFs in \figSS{partial_pdf}). In these simulations only the noted pair varies change its fraction (x axis) while the other set of pairs remains constant and equal to 1. We see that whenever clustering occur, meaning that the fraction of Zr--X decrease with respect to the other set of pairs (blue curve),  $r^{1}_{max}+r^{2}_{max}$ decrease as well. The blue shadow  projection represent the region of values that are experimentally relevant (cf. color bar in (b)). We note that changes Ni--Ni fraction,for example close to the Ni-BMG interface, are benign. Changes in Zr--Zr fraction, for example in possible Zr-rich parts of the sample may effectively reduce the $r^{1}_{max}+r^{2}_{max}$ parameter. However, as shown in \figSS{r1r2eels}, features in the $r^{1}_{max}+r^{2}_{max}$ map do not resemble features in a [Zr]/[X] concentration map taken from the same area, and suggests that the effect of Zr concentration on the intensities in map (b) in not dominant.}
 \label{fig:structural}
\end{figure}

We now attempt to find QoIs that emphasize specific structural features of interest. One of them would be the overall bond distributions. Assuming that the static disorder in a glass is greater than any thermal disorder, a peak width will represent the static distribution of bond distances weighted by the atomic number. In regions where the average concentration is not varying too much, the spatial variation of the height of a given PDF peak is a reasonable indicator of the bond-distribution, with lower peaks indicating a broader distribution. Comparing $G(r^{1}_{max})$ and $\sigma^{1}$ maps (\figs{gauss}(d,f)), we find that mostly this rule is being followed. A QoI that is the ratio between $G(r^{1}_{max})$ and $\sigma^{1}$, can be expected to be an even more sensitive QoI for the bond distribution, and this is plotted in \fig{structural}(a), which represents the peak sharpness, with a darker color indicating a broader bond distribution.

We now return to the issue of chemical short-range order.  We note that, if there is a tendency towards phase separation in an [A,B] mixture, A-A and B-B pairs are over-represented and A-B pairs are under-represented compared to the value for the random packing.  On the contrary, if there is a tendency for chemical ordering, as would be the case in a compound formation, then A-B pairs will be over-represented and A-A and B-B pairs under-represented compared to the average. We can consider this type of analysis for the BMG in the central region (away from the Ni-rich edges) because the composition is overall uniform and expected to be close to the nominal \break \thebmg composition.  As before, we treat the BMG as a pseudo-binary Zr/X compound.   Similar to the fitting effort done for \fig{coordshell}(c) from which we extracted a QoI that represents chemical SRO, we would like to have a measure of how the ratio of Zr-X varies with respect to Zr-Zr and X-X as a function of position. Without dealing with the cumbersome process of fitting realistic structure models, we would like to extract a QoI from the simple two-Gaussian fits that will be sensitive to chemical SRO. We are fortunate that in the PDF Zr-X lies to the left of the first CS doublet and contributes to Gaussian 1, and Zr-Zr lies to the right and contributes to Gaussian 2.

In a search for QoIs that are most sensitive to the chemical SRO for a given average composition, we tested a number of possible QoI's, such as $r^{2}_{max}-r^{1}_{max}$ and $r^{1}_{max}+r^{2}_{max}$. With the reasoning that the X--X peak is the shortest in Gaussian 1 and Zr--Zr is the shortest in Gaussian 2, chemical clustering may result in reduction in the value of the summed Gaussian centers, i.e., reduction of $r^{1}_{max}+r^{2}_{max}$. To confront this hypothesis, we have carried out a simulation of the summed peak positions as a function of the relative ratio of each pair, at constant overall composition. In \fig{structural}(c) we show the resulting $r^{1}_{max}+r^{2}_{max}$ values when changing the relative ratio of each pair in the simulated data. We see that $r^{1}_{max}+r^{2}_{max}$ does get smaller with clustering as suggested by our assumption. Overall, we conclude that the $r^{1}_{max}+r^{2}_{max}$ QoI reflects the change in the chemical SRO. Later in \figs{nmf}(e) and (f) we cross validate this result using NMF analysis.

%

\subsection{Mapping global similarities and principle component reconstruction}

\begin{figure}
	\centering
 	\includegraphics[width=0.50\linewidth]{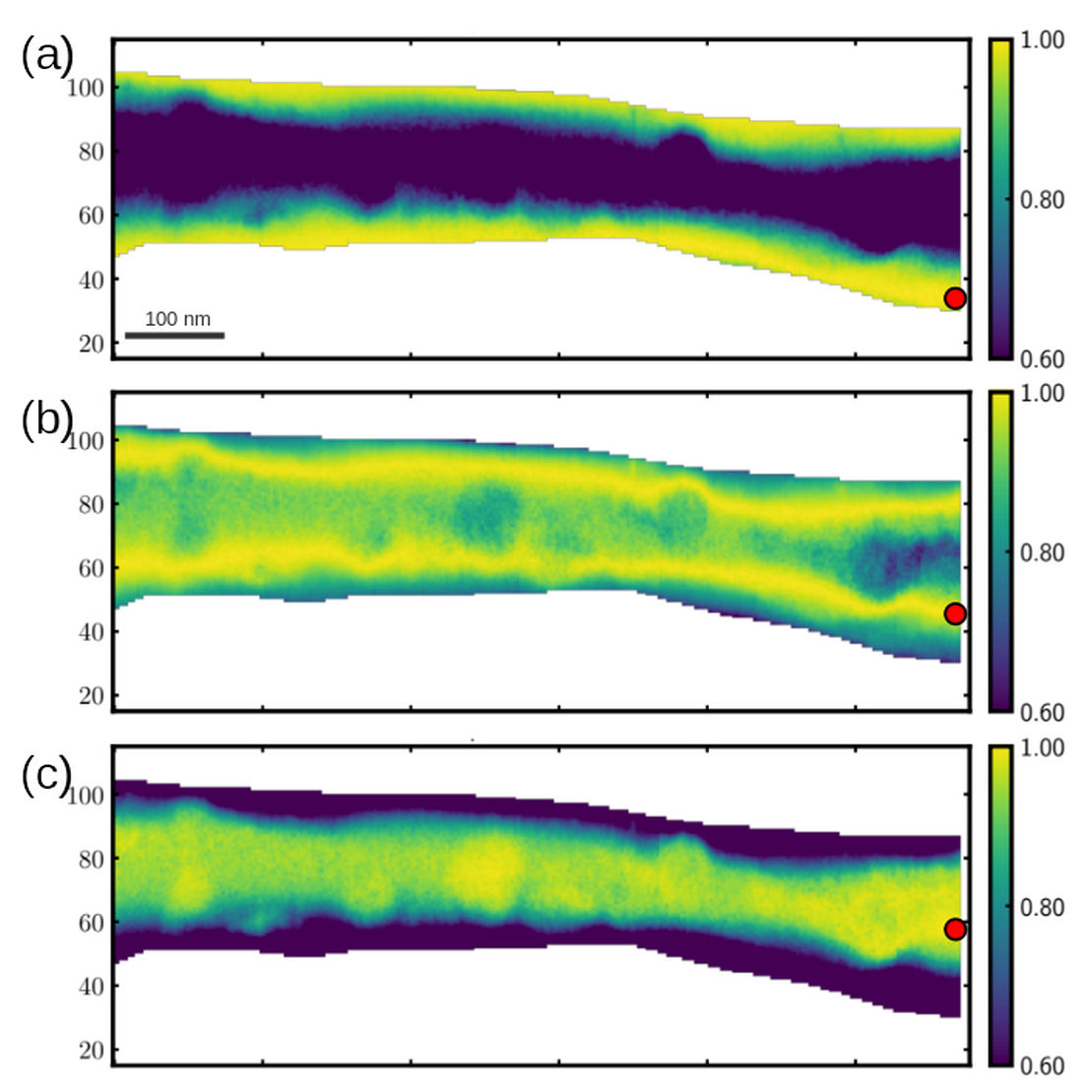}
 \captionof{figure}{Similarity maps using the Pearson correlation coefficient as QoI. The Pearson correlation analysis extracted a coefficient using a selected pixel (within the red dot on each map - see right-hand-side of the maps) and comparing it with the rest of the unmasked dataset. Masked areas are colorless. The x-y range of the maps are the same as for the other SNEM maps throughout the paper. Each map (a)-(c) represent a major subgroup of ePDFs across the dataset. Here we mask data that refer to the crystalline-Ni regions, which emphasizes the BMG region.}
 \label{fig:pearson}
\end{figure}

Up to this point we focused on a selected region or cut that was identified and rationalized based on prior knowledge about the sample. In some cases such information is absent, and even if present, it is simpler to classify the data into sub groups based on similarity. Using Pearson correlation analysis~\cite{jensen_x-ray_2015} between a selected ePDF and the other ePDFs, where from each comparison we derived the Pearson coefficient as a QoI, one can create Pearson similarity maps as we have done in \fig{pearson}.

In this approach a pixel is selected and the similarity of the PDF in all other pixels to that in the selected pixel is computed and a map plotted.  The Pearson correlation coefficient takes a value of 1 if two curves are perfectly correlated, 0 if they are uncorrelated, and -1 if they are anti-correlated, and it can take any value in between these limits depending on the degree of correlation.  It allows the experimenter to choose a region of the image that appears interesting, and find all the other regions that are similar.  It has also been used in a powerful way to pick pixels that are not interesting (for example, they come from a background signal such as a host lattice) and to find all the pixels that deviate from that~\cite{jense;m21}. To demonstrate we have made three Pearson maps that are in \fig{pearson}. Panel (a) represent ePDFs in the region that is close to the Ni-BMG interface, panel (c) to that of the BMG core, and panel (b) is an intermediate that links between region (a) and (c). Following the discussions we had so far, we can immediately associate panel (a) with a Ni-rich set of ePDFs, the Zr-rich region is represented by the panel (c), and a characteristic transition region between the two Ni and Zr-rich regions is characterized my panel (b).
This interface region shown in (b) refers, most-likely, to a region with a particular concentration range that impose a particular characteristic structure. We can see that a similar trajectory is shown as a white line in the [Zr]/[X] concentration ratio map that is shown in \figSS{r1r2eels}(b), where the white line represents the region where [Zr]$\approx$[X].
To focus our interest on the BMG, we masked the crystalline Ni region based on the other maps (specifically \fig{structural}(b)) and applied the Pearson correlation analysis on the unmasked ePDFs. For Pearson analysis masking is not a requirement but only assists in visualizing differences within a particular region of interest.

Next, we introduce and use non-negative matrix factorization on an ePDF dataset that includes the same ePDFs shown in \fig{pearson} (after masking the Ni region.\footnote{We found that masking the ``outlier" pixels containing noise that were associated with the crystalline nickel regions was important to get a good NMF decomposition}), and extract components that are physically-significant, along with the weight matrix~\cite{liu_validation_2021}. These can then be used as an important set of QoIs.

A recent work~\cite{mu_unveiling_2021} demonstrated the power of independent component analysis (ICA) for classifying relational ePDFs from heterogeneous amorphous systems to their weighted partial contributions. NMF shares the same goal as the ICA algorithm, and a successful reconstruction of x-ray PDFs from their weighted partial contributions has been demonstrated~\cite{liu_validation_2021}. Independent to the ICA success, below we demonstrate that NMF also works exceptionally well for reconstructing partial ePDFs from spatially resolved sets of ePDF data.

Like Pearson analysis, NMF is useful for an exploratory analysis when one does not have prior knowledge on the chemistry or structure of a system~\cite{liu_validation_2021}.
NMF is related to principle component analysis but it seems to produce (mathematical) components that explain the variability of the data in the set of PDFs that are more physically meaningful~\cite{lee_learning_1999}, and in particular of interest here, when applied to PDF data~\cite{liu_validation_2021}.  We note that PDF data goes negative, and therefore requires shifting to positive values for this procedure. We found that it worked better when shifted to positive values well above the minimum value in the PDF. The analysis was carried out using the python-based scikit-learn~\cite{pedregosa_scikit-learn_2011} NMF decomposition package. Since the number of components is a required input, we found that four components has a sufficiently low reconstruction error and that it returns a desired set of physically-relevant NMF components as explained below.

\begin{figure}
	\centering
 		\includegraphics[width=0.95\linewidth]{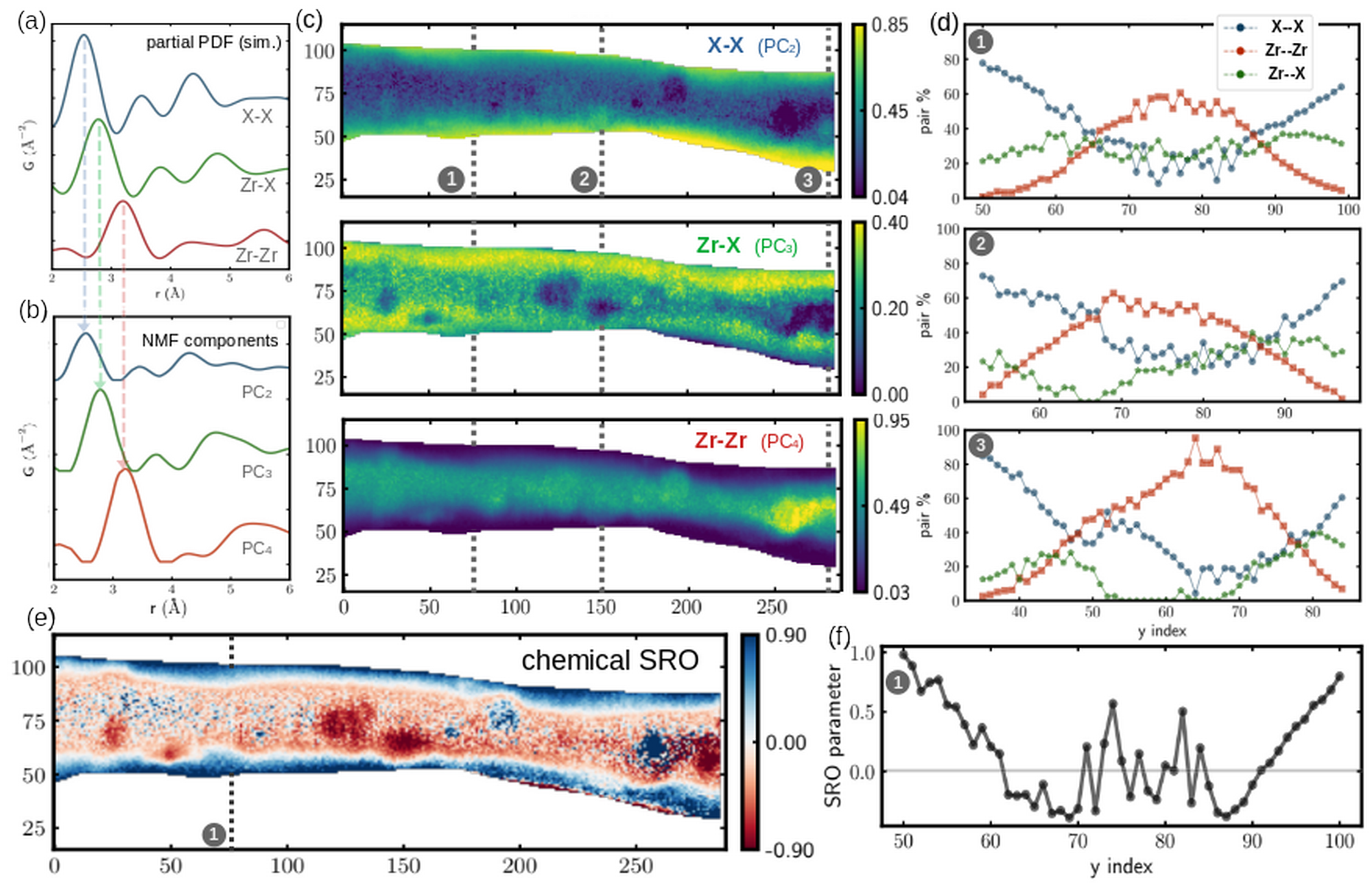}
 \captionof{figure}{NMF analysis using the entire set of ePDFs (excluding the Ni masked regions - same as for \fig{pearson}) as the input matrix. In (a) we show simulated partial PDFs for closed-packed X--X, Zr--X and Zr--Zr pairs and compare them to the 2nd-4th factorized principle components (PC). This comparison emphasize the similarity and thus the relevance of the NMF components to the actual pair density. In (c) we map the weights of each NMF principle component into a QoI map. In (d) we plot selected cuts across the y dimension axis to show the interplay between the pair densities, similar to what was plotted in \fig{coordshell}(b). In (e) we map the chemical SRO parameter, $\alpha$, following \eq{sro} and \eq{p_max}. In (f) we plot a profile along the y dimension axis of the selected cuts 1, similar to the plotted SRO parameter in \fig{coordshell}(c).}
 \label{fig:nmf}
\end{figure}

The NMF decomposition of the PDFs from the BMG region resulted in the top four components shown in \fig{nmf}(a). Indeed we find a striking resemblance in three out of the four components that are plotted in \fig{nmf}(b) to the partial PDFs of Zr--X, Zr-Zr and X--X (\fig{nmf}(a)).
The resemblance between the simulated and the NMF components (i.e., panels (a) and (b) in \fig{nmf}) is remarkable mostly because the NMF analysis suggested a set of realistic local structure components (in the form of partial PDFs) without any scientific intuition or prior knowledge. Without diving into the detailed analysis of the remaining component, since it scales with the amount of the positive shifting we assume it is a background intensity that is common for the entire dataset. This is unlike the other three components that remain fairly unchanged when scaling the shifting.

Since NMF returns a weight matrix that contains the contribution of each component to the approximated experimental ePDF, we can use each weight value as a QoI that approximates the partial density of each pair. These QoIs are normalized to sum up to 100\% which are then used to create the `bond' maps shown in \fig{nmf}(c). We clearly identify the regions that are X-rich close to the Ni-BMG interface and those that are Zr-rich at the midline of the BMG. The Zr--X density from \fig{nmf}(c) (middle map) does not show a clear tendency with the Ni- or Zr-rich regions. Zr--X bonds seem to be deficient in a few of the regions that contain nano-crystallites (cf. \fig{VDF}), while enriched in a particular [Zr]/[X] concentration window along the Ni-BMG-Ni cut (cf. \figSS{r1r2eels}(b)).

To investigate the bond-type evolution across the Ni-BMG-Ni profile we chose three cuts through regions of the sample that are visually different, noted as 1-3 in \fig{nmf}(c). \fig{nmf}(d) shows the interchanging evolution in pair density of each pair. The obvious variation between the profiles emphasizes the richness in the distribution of local order, especially along the transition between the BMG-core to the Ni-BMG interface. Note that panel (1) in \fig{nmf}(d) is representing the same cut that is used for the model-fitted ePDFs in \fig{coordshell}(b). One can appreciate the similarity between the two in terms of the trends. However, the absolute values of the assigned pairs is somewhat different, especially the magnitude of the Zr--X and Zr--Zr pair densities. One should remember that NMF decomposition still returns mathematical components that are not guaranteed to fulfil physical constrains, thus should be treated as semi-qualitative, unless proven otherwise. On the other hand, the model-fitting uses a perfect closed-packed FCC model, which is an approximation, where a more complex non-perfect packed model would have resulted in a better agreement with the NMF result. Future work for benchmarking and calibrating NMF components in SNEM against a structural model with a known and controlled dataset is thus needed.

%

Having values for the partial pair densities, we can now, using \eq{sro} and \eq{p_max}, generate a full chemical SRO map (\fig{nmf}(e)) and a cross-cut profile (\fig{nmf}(f)) to compare with \fig{coordshell}(c) that was generated from the the model-fitting. Since the chemical SRO should oscillate around zero (the random distribution case), the white-colored regions in \fig{nmf}(e) represent a random distribution, red coloured regions are those that exhibit chemical clustering, while blue regions represent regions with enhanced Zr--X ordering. The map shown in \fig{nmf}(e) shows a remarkable insight to the distribution of the local order in a heterogeneous glassy material. This should help to answer questions that relate to, for example, the mechanical, electrical and magnetic properties of glasses, such as nucleation of shear-bands, propagation of defects, electrical conductivity and formation of magnetic centers, as a function of the local structural and chemical order in the material.

The resemblance between \fig{nmf}(f) and \fig{coordshell}(c) is clear, showing oscillations in the SRO parameter around zero with negative and positive values of the SRO parameter around the same regions. This once again validates the power of NMF as a semi-quantitative approach that can purely mathematically extract structurally-relevant features and their partial contribution from a relational dataset, related in time and/or space.

\subsection{Relation between order and composition}

\begin{figure}
	\centering
 		\includegraphics[width=0.80\linewidth]{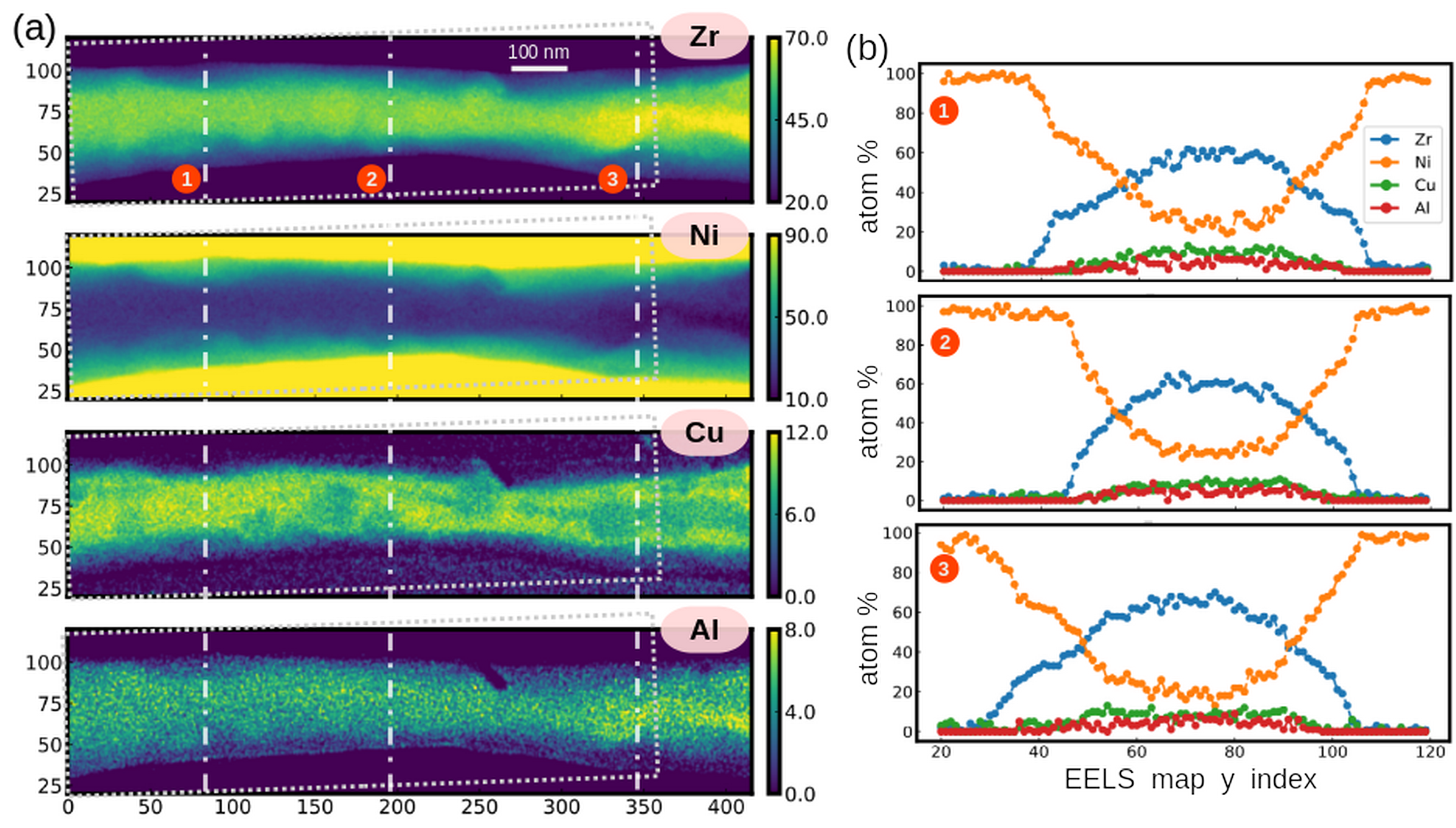}
 \captionof{figure}{(a) EELS composition maps of (top to bottom) Zr, Ni  Cu and Al. The values in the color bars represent the partial percentage, which were extracted using the L-edge-Zr, L-edge-Ni, L-edge-Cu, and K-edge-Al. The dotted frames represents the approximate region of the SNEM maps that are presented along the paper. In attempt to emphasizing regions of contrast, a Gaussian smoothing filter is applied. (b) Selected concentration profiles of all four elements taken from cuts 1-3 (see dashed markers in (a)). These selected cuts correspond to those presented in \fig{nmf}(d) for comparison.}
 \label{fig:eels}
\end{figure}

To this point we pointed out the chemical composition as a factor that should correlate with the pair density. While SNEM focuses on structural features, or the way chemical species are coordinated among themselves, the chemical composition provide information about what are the used building-blocks in a structure. Although these two terms are related, they project different information that is not necessarily correlated.  Augmenting compositional data on-top of SNEM, i.e. structural information, should be of value to address the question of structure-composition relation. In \fig{eels} we present EELS composition maps and select cut profiles. The selected cuts in \fig{eels}(b) correspond to those in \fig{nmf}(d) for the sake of comparison between structural and compositional spatial evolution in the structure.

%

From \fig{eels} we learn that as we get closer to the Ni-BMG interface, and as earlier concluded from the SNEM results, the BMG becomes gradually Ni-rich at the expense of the other elements, especially Zr. We find that the X compound is effectively Ni, with little contribution from Cu and even less from Al.
This is in contrast to the nominal BMG concentration where Ni and Cu should occupy 10\% and 17.5\%, respectively. The Ni rich region at the interface was likely to be possible due to the surrounding crystalline Ni amorphizing during the hot-rolling fabrication process.

In a comparison between the atomic and pair distribution maps (i.e., \fig{eels}(a) with \fig{nmf}(c)) and profiles (i.e., \fig{eels}(b) with \fig{nmf}(d)) we find that both suggest Ni enrichment closer to the edge of the BMG, but a closer look shows that these profiles are not identical. The main feature that the composition maps cannot predict is the partial density of Zr--X pairs. The main feature that SNEM cannot provide is the density of each atom in space. The augmentation of the two parts to a unified picture is therefore very powerful as
will be discussed below.

Unlike the BMG edge effects that can be mainly explained by Ni enrichment, the different SNEM maps exposed inclusions within the core of the BMG. Similar inclusions and also in similar locations are clearly seen in the Cu EELS map as Cu depleted regions. A quick visual correlation between the Cu-depleted regions with the SNEM maps suggest that Cu deficiency correlates with nucleation of nano crystallites (cf. \fig{VDF}).  Either nucleation is preferred in regions that are deficient in Cu, or Cu solubility in the nucleating species is low and the Cu is excluded from the nucleated region.  To the best of our knowledge, such correlation between Cu-deficiency and the mentioned structural and mechanistic correlations was never reported before.

At this point it is not very clear what is the physical or chemical source for the impact of the Cu-deficiency on the structure. It is known~\cite{lee_crystallization_2016} that 1:1 ZrCu glasses tend to crystallize slightly above the glass phase transition, namely at $\sim 700^{\circ}$~C. Since the critical crystallization temperature in metallic glasses tends to decrease with a decrease in the degree of mixing, and since we performed the rolling process at elevated temperatures ($420^{\circ}$~C), small local variations in composition might  have resulted in nucleation of nano-crystallites. 

\rev{2.11}{As for the Ni-BMG interface, it is unsurprising that Ni and Zr intermix  across the BMG-Ni interface. The roll bonding at 420$^{\circ}$C occurs between the glass transition temperature (361$^{\circ}$C) and the crystallization temperature (466$^{\circ}$C) where both thermal and shear-driven inter-diffusion, might be expected.}


At this point these are merely hypotheses that require further study. Nevertheless, our observations provoke a solid direction for study that could not be justified without the SNEM experiment. Visualizing and understanding the chemistry-structure-process relations have immediate implication on understanding formation of shear-bands in metallic glasses. The link between chemistry and structural order is an important tier to the ongoing studies in metallic glasses~\cite{wu_atomistic_2021,yang_susceptibility_2019, sopu_atomic-level_2017, takeuchi_atomistic_2011, ye_atomistic_2010}, but also in other fields, such as structure evolution in batteries~\cite{christensen_understanding_2021, yang_atomic_2019}, and biomineralization~\cite{de_yoreo_-situ_2016}.

\section{Conclusion}

This paper is the first in a series to show imaging of quantitative structural evolution in heterogeneous systems at a nano-metric length scale. Inspired by previous works done at synchrotron facilities~\cite{kovyakh_towards_2021, wright_towards_2020}, in this work we develop the use of spatially resolved ePDF, SNEM, for mapping fluctuations of the local structure, but with a spatial resolutions that is several orders of magnitude higher than what can be achieved in a synchrotron. 

This extension to the 4D-STEM methods, where order evolution obtained from quantitative ePDFs is followed, is a powerful approach to elicit fine details about local structure in amorphous and nanocrytsalline regions of a sample. We demonstrate that different aspects of the local structure can be revealed by defining different quantities of interest, which can be very trivial to extract (\fig{overview}), or may require the use of more detailed structural models (\fig{coordshell} and \fig{structural}). We demonstrate how the large spatially resolved dataset opens the door to using statistical and machine learning tools such as Pearson correlation analysis (\fig{pearson}) and NMF  (\fig{nmf}) to automatically cluster and group regions with similar structural signature, but also to disentangle a signal coming from complex structure into principle physically-meaningful components of partial structural motifs.

Using SNEM, we addressed the question of how local order evolves in metallic glasses when encapsulated in a crystalline matrix. In particular, we were able infer compositional variations and find regions with chemical short range order or clustering (\fig{nmf}(e)). This could be combined with additional chemical information from EELS (\fig{eels}) and nano-crystallization from virtual dark field images (\fig{VDF}) to gain insights into the nature of chemical processes that the BMG underwent during hot rolling. We demonstrated that regions that are Cu-depleted correlate with sites where nucleation and growth of nano-crystallites has occurred. Moreover, we found significant heterogeneity in the chemical ordering of atoms around themselves, that were pronounced around nucleation sites and next to highly-deformed region at the BMG-Ni interface. 

\rev{2.4}{Since local ordering define the local chemical potential and bond-nature, it must affect the global properties of materials. And, indeed, recent works~\cite{wang_effects_2020, louzguine-luzgin_shear-induced_2021} have demonstrated that microscopic ordering in glasses affect mechanical and corrosion properties, and that stress can impact the local chemical order. These are a few examples where high resolution local structural information can be augmented using SNEM, as it carry the potential to expose causality relations of order to composition and properties.} 

SNEM measurements are quite straightforward and do not require very specialized instrumentation. Using the SNEM methodology, we show that integration of both simple peak-finders and more complex machine-learning classifiers unlock hidden compositional and structural information. However, it is important to acknowledge that the implementation of the SNEM approach would not be possible without the integration of numerous experimental and data analytical tools, such as scanning precession electron diffraction, rapid PDF acquisition and analysis infrastructure that were originally developed for synchrotron x-ray diffraction, along with many open-source computational tools that could be woven into coherent data-analytic pipelines.

It is also important to acknowledge the resolution SNEM provides, both spatial, as demonstrated in this work, but also temporal. The nm spatial resolution is found to be critical for imaging structural evolution, and thus understanding its correlation with composition and applied stress during fabrication.
With the growth of \textit{operando} experiment setups~\cite{nakamura_situ_2020, xu_chemistry_2019, morandeau_situ_2015},  we predict a fruitful integration of SNEM in regular material discovery when temporal resolution is required. Besides the fact that electron diffraction for SNEM requires generally lower doses with respect to other elemental analysis and spectroscopic techniques, if a system evolves during the time of the experiment, different QoI's coming from the rapidly acquired diffraction patterns (here $0.1$~sec per frame) may allow to trace temporal structural evolution.  Rapid synchrotron acquisition and analysis of x-ray diffraction and PDF analysis with time resolution $<1$~sec has already been demonstrated~\cite{wright_towards_2020, chupas_rapid-acquisition_2003}. Thus, the current SNEM infrastructure, meaning integration of rapid acquisition and analysis tools, has the potential to follow processes that happen under the microscope and track very fast structural changes in the most complete manner known today.



We find SNEM as a strong candidate for taking the lead in data-driven exploration of local order in materials in advanced metallurgy, energy materials~\cite{christensen_understanding_2021, yang_atomic_2019}, bio-mineralization~\cite{de_yoreo_-situ_2016}, and more. As for classical materials, SNEM will allow us to ask new questions on more classical materials that are yet to be answered and relate to the local order and its evolution in time and space.

While it may contribute alone in these fields of research, we note that SNEM is complimentary to other electron spectroscopy based techniques. We would like particularly to emphasize EXELFS~\cite{williams_fine_2009, hart_synchrotron_2019}. Unlike SNEM, EXELFS provide local structural information that is highly complementary to SNEM and yields the very local environment around particular target atoms. While SNEM require no energy-filters, provide effectively higher spatial resolution and require smaller doses, moving forward we expect to see a complementary development and application of these methods. Although each method is important on its own, as demonstrated with the augmentation of compositional mapping with EELS (\fig{eels}), a collaborative work between SNEM and EXELFS should increase the flexibility and richness in collecting spatially-resolved local order information in materials.

\label{sec:conclusions}
\section{Acknowledgements}
J.L.H and P.P.D contributed equally. Y.R. thank Songsheng Tau for his assistance in developing the MiniPipes code.
Development of data analysis pipelines and protocols in the Billinge group was funded by the Next Generation Synthesis Center (GENESIS), an Energy Frontier Research Center funded by the U.S. Department of Energy, Office of Science, Basic Energy Sciences under Award Number DE-SC0019212.
Electron microscopy in the Taheri group was supported in part from the U.S. Office of Naval Research through contracts N000142012368 (a Multidisciplinary University Research Initiative (MURI) program) and N000142012788, and in part from U.S. Department of Energy, Basic Energy Sciences, through contract DE-SC0020314.
Sample creation and preparation in the Mathaudhu group was funded by the U.S. National Science Foundation under CMMI Grant 1550986.
\label{sec:acknowledgements}

\bibliographystyle{elsarticle-num}
\bibliography{20yr_snem}

\clearpage
\setcounter{figure}{0}
\setcounter{table}{0}
\setcounter{section}{0}
\makeatletter
\renewcommand{\fnum@figure}{Fig.~S\thefigure}
\renewcommand{\fnum@table}{Table~S-\thetable}
\makeatother

\begin{center}
\Huge\textbf{Supplementary Information}
\horizline
\Large{\mytitle}
\horizline
\end{center}
\begin{figure}[h]
 \centering
 		\includegraphics[width=0.65\linewidth, valign=t]{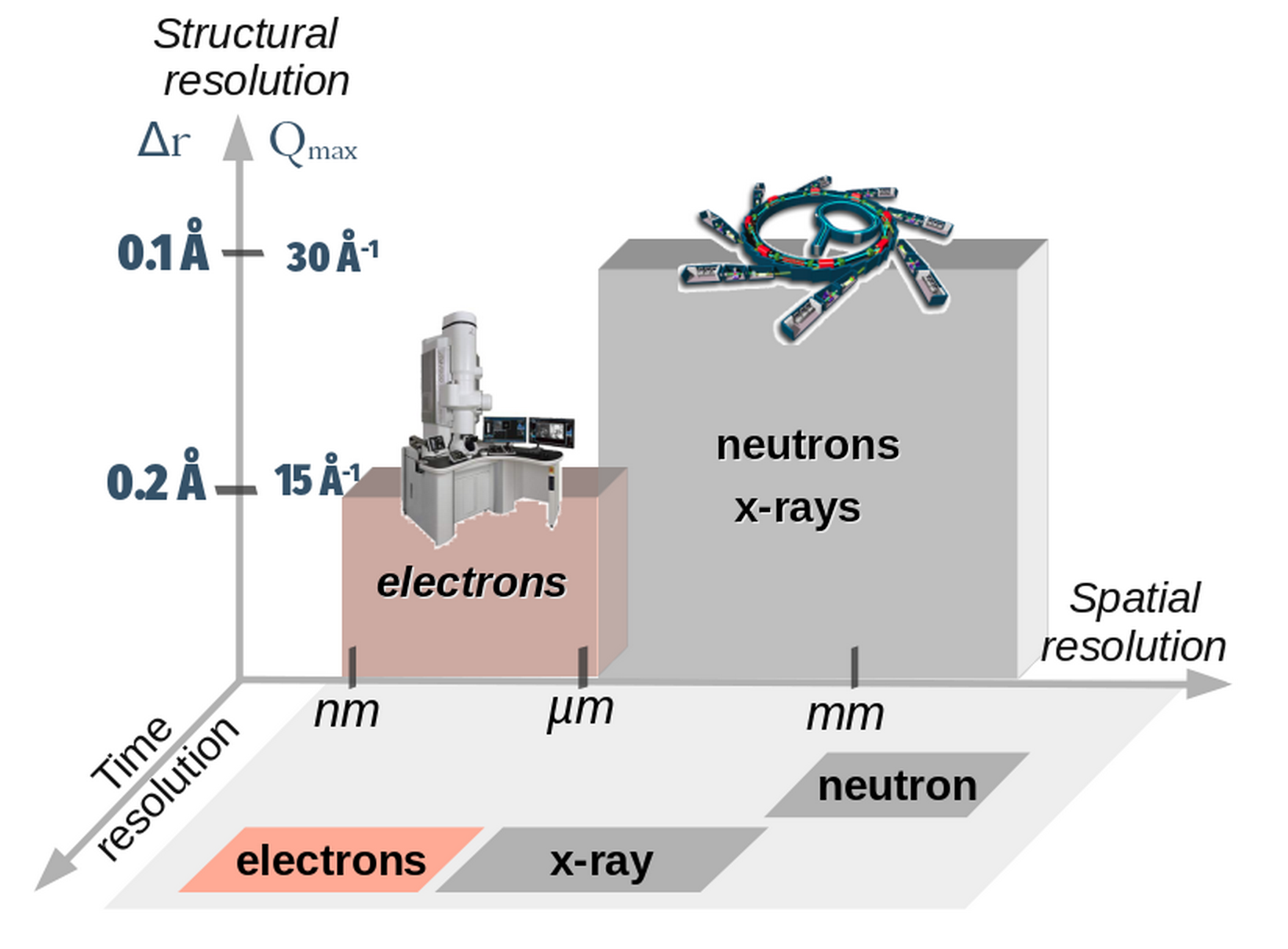}
 \caption{ PDF resolution roadmap in high energy x-ray, neutron and electron total scattering experiments. Electron diffraction offers the best positional resolution (smallest beam size), while x-ray and neutron diffraction offer the best structural resolution (higher maximum collected momentum transfer value, $Q_{max}$, which, based on the Nyquist-Shannon sampling theorem, allows a better r-space resolution ($\Delta r=\frac{\pi}{Q_{max}}$)~\cite{farrow_nyquist-shannon_2011}).
Neutron acquisition time is slow and beam size is large ($>$ mm), but it can offer better elemental contrast and additional insight on chemical and magnetic order~\cite{mirebeau_diffuse_2017}. The time resolution for electron or x-ray sources depends on the flux, the desired $Q_{max}$, and the chemical elements under study. Using fast detectors, rapid acquisition and automated analysis pipelines, PDFs can in principle be generated in a fraction of a second \cite{chupas_rapid-acquisition_2003, juhas_pdfgetx3_2013} which makes it suitable for \textit{operando} experiments.}
 \label{fig:resolutions}
\end{figure}

\begin{figure}[h]
 \centering
 		\includegraphics[width=0.7\linewidth]{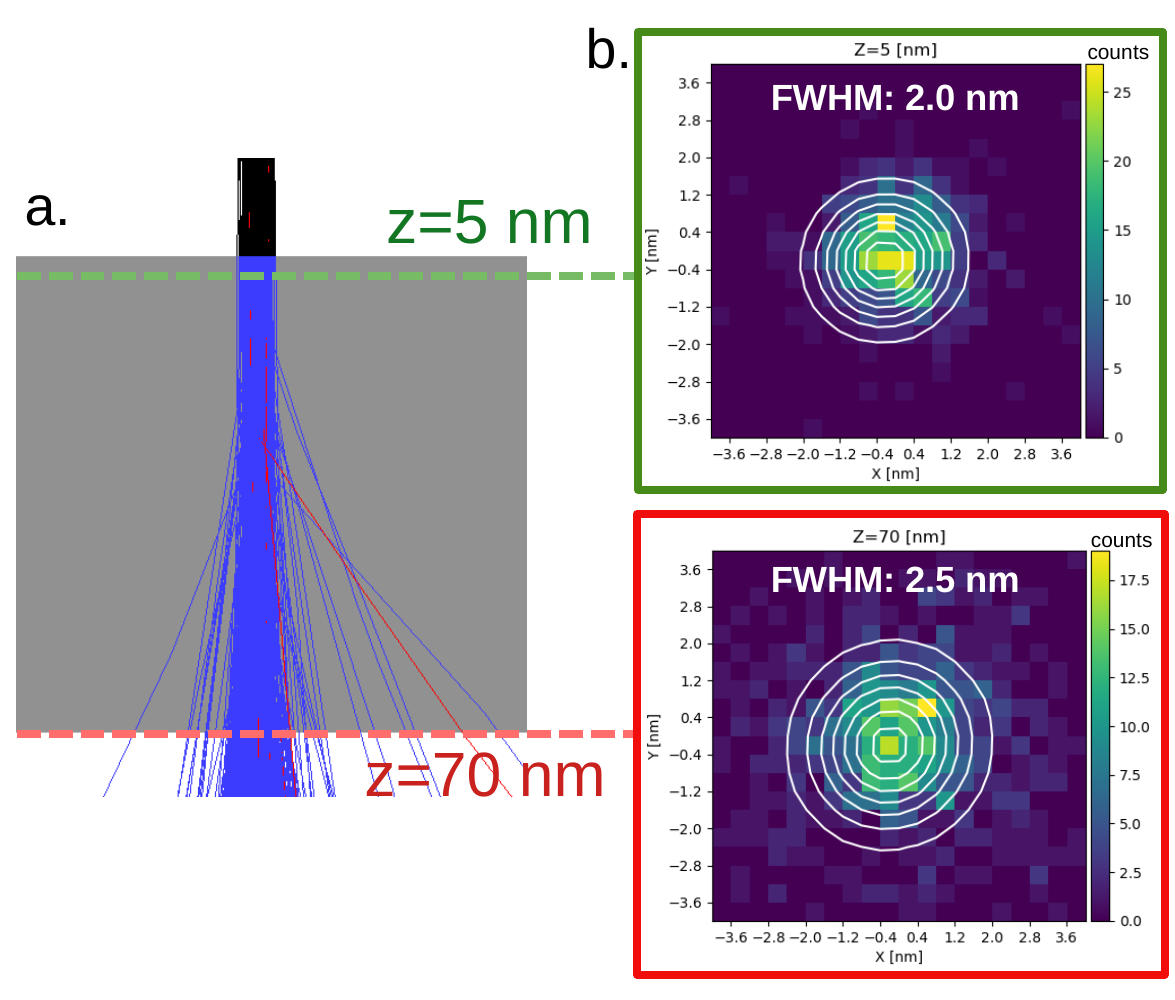}
 \caption{(a) Monte Carlo simulation using CASINO~\cite{hovington_casino_1997}. The simulation shows the trajectory of 2000 electrons coming from a 2 nm wide 200 keV electron beam with a convergence angle of 1 deg (17 mrad) (following the 1 deg precession angle) passing through a 70 nm medium composed of $\mathrm{Zr_{60}Ni_{25}Cu_{10}Al_{5}}$. (b) the intensity projection at z=5~nm and z=70~nm. The FWHM at the exit of the sample (z=70~nm) is $\sim$ 0.5~nm wider than at the entrance to the sample (z=5~nm).  For the simulation we used 2000 electrons and a virtual detector bin size of $0.4\times0.4$~nm}
 \label{fig:casino}
\end{figure}

\section*{Methods}\label{sec:methods}
\section{Sample preparation} \label{sec:sample_preparation}
The sample under study is a roll-bonded laminate of nickel and a bulk metallic glass (BMG) formed as the result of accumulative hot rolling. The preparation of the sample was discussed in detail in Refs~\cite{hart_synchrotron_2019, shahrezaei_synthesis_2019}. Briefly, sheets of Ni (99.99\%) and the BMG  material (\thebmg at.\%), with initial layer thicknesses of 0.2~mm and 0.06~mm, respectively, were placed in a vertical stack. The stacks were heated in Ar to 420~$^{\circ}$C and rolled to 50\% reduction using a heated two-high rolling mill (IRM model 4060). Following rolling, the sample was cooled to room temperature and cut in half with a mechanical saw. The two halves were then stacked vertically and spot welded together, and then the rolling process was repeated. In total, six rolling steps were used to form the accumulative roll bonded laminate. For the SNEM and EELS analysis, a focused ion beam (FIB) was used to prepare a $70\pm15$~nm  lift-out. Final thinning was performed with 5~keV Ga ions. The final sample-region that was scanned, the sample region-of-interest (RoI), - a single Ni/BMG/Ni laminate - is shown in \fig{intro}(b).

\section{SNEM data collection and normalization} \label{sec:data_collection}
The goal of the data acquisition for SNEM imaging is to obtain ED data that will result in good ePDF's~\cite{egami2012underneath, abeykoon_quantitative_2012, gorelik_total-scattering_2015, hoque_structural_2019, khouchaf_study_2020}, i.e., data collected over as wide a range of reciprocal-space as possible with high signal-to-noise ratio, good statistics and as kinematical as possible~\cite{claffey_kinematical_1970}, whilst minimizing potential beam damage, sample charging and other effects that will degrade the signal integrity~\cite{gorelik_total-scattering_2015}. With this in mind, the following instrument setup conditions were used.

\subsection{diffraction pattern generation} \label{sec:diffraction_pattern_generation}
A JEOL 2100F microscope was used in a STEM mode with a Schottky electron source, operated at an accelerating voltage of 200~kV. The beam scanning and precession is controlled by an external ``Digistar'' device~\cite{nanomegas, rauch_automated_2010, rauch_automated_2011}. We used the smallest (10~$\mu$m) condenser aperture, which allowed a beam spot size of $\sim 2$~nm; the raster step-size was chosen to be 3~nm. The diffraction length was set to 10~cm, and the overall TEM image magnification was 8000~X. We used a precession electron diffraction mode~\cite{vincent_double_1994}, where each pixel was collected with 8~precessions per cycle at a precession angle of 1$^{\circ}$. With a dwell time of 0.08~s per pixel, collection from a $289\times 132$ grid (in total: 38,148~diffraction patterns) took 51~min and 25~s. Given the long acquisition time, an occasional drift correction from a reference area outside the sample RoI was carried out.
To minimize potential beam damage, we used the smallest condenser aperture and minimized the dwell time. No sample degradation was observed.

The patterned images of the diffracted electrons were acquired by an $8.3 \mu\mathrm{m} \times 8.3 \mu\mathrm{m}$ pixel-size CCD Stingray camera (Model No F046B ASG) that was placed outside the microscope, capturing the fluorescent screen image from above. Although the camera allowed a 14~bit dynamic range, to accelerate data acquisition with a sufficient signal to noise ratio, we used an 8-bit dynamic-range setting and $4 \times 4$~binning. The 8-bit dynamic-range was sufficient to cover the observed intensities and saturation was observed only from the direct beam and some strong Bragg peaks coming from the crystalline Ni. With a total of $576 \times 576$~pixels and $4 \times 4$~binning, each captured DP included $144 \times 144$~pixels

\subsection{Diffraction pattern distortion correction} \label{sec:distortion_correction}
Besides experimental setup factors, including the external inclined position of the camera (see \fig{overview}(a)), geometric distortions to the diffraction pattern images are naturally introduced. A distortion correction was carried out using the ``ASTAR" software package~\cite{rauch_improving_2010, portillo_precession_2010}, which resulted in a concentric circularly symmetric DP centered at the middle of the $144 \times 144$ square image. The distortion-correction process uses a de-warping numerical process that involves matching between an experimental DP from the crystalline Ni region and the expected peak positions from the relevant Ni orientation.
Choosing a DP from a particular Ni crystallite that is oriented close to the [102] zone-axis, allowed us to increase the certainty of the fitting. The identification of the approximate [102]~Ni pattern within the capping Ni region of the sample followed the description  in the supplementary information \sect{orientation_match}.
Eventually, the stretching factors were extracted from matching the DP at the x=26, y=27 pixel (\figSS{calib}(a)) to a Ni-[102] simulated DP. To normalize the entire dataset, the extracted stretching factors were applied on all images in the dataset. The corrected DPs were stored and transferred for QoI-extraction via our analysis pipeline. Selected post-normalization DPs are shown in \fig{overview}(b).

\subsubsection{Orientation match and matching reliability maps}\label{sec:orientation_match}

This section describes briefly how phase and orientation matching and related reliability index map were generated using ``ASTAR'' software package and how they are used for generation of stretching parameters for distortion correction of DP's before SNEM data analysis~\cite{rauch_automated_2010}.

\begin{figure}[h]
 \centering
 		\includegraphics[width=0.8\linewidth, valign=t]{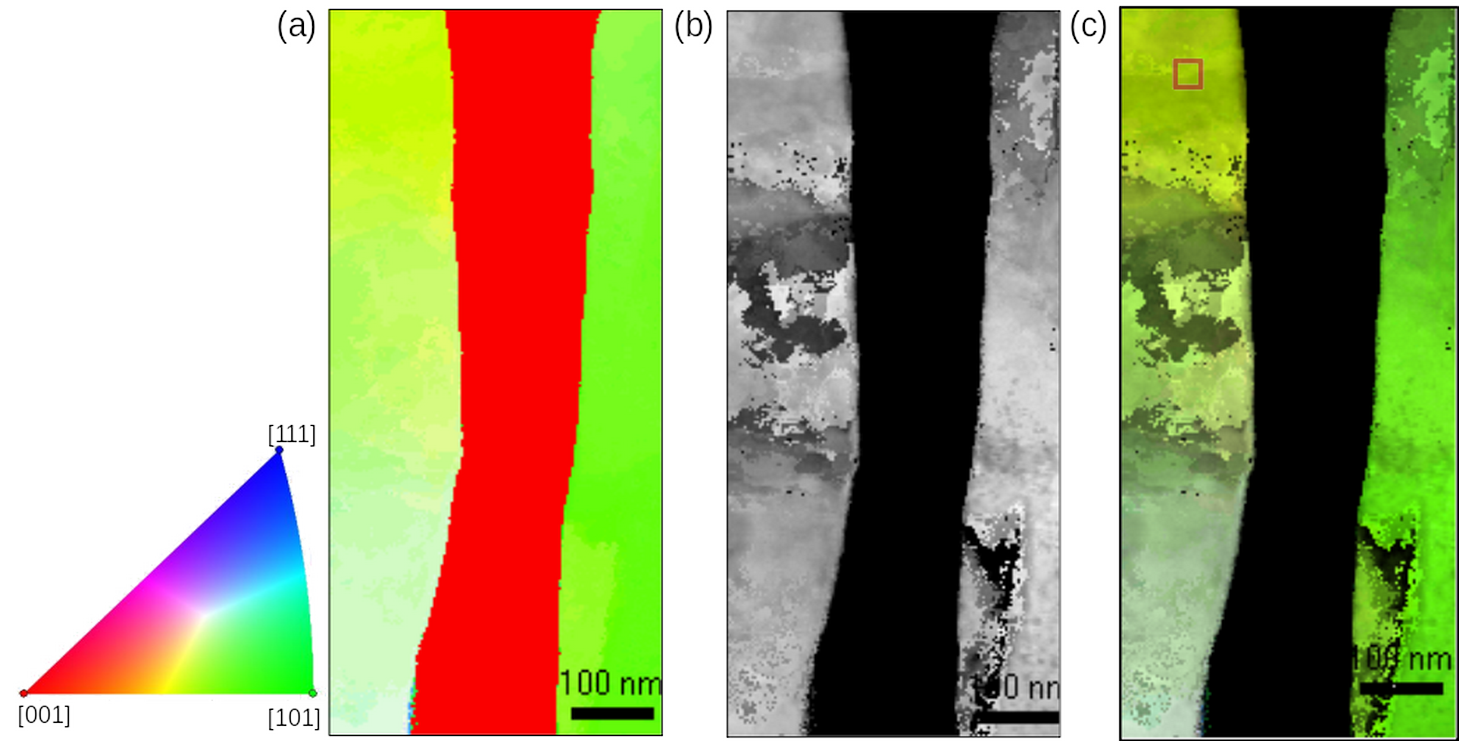}
 \caption{(a) Orientation matching map and (b) the corresponding orientation reliability map as resulted from the distortion corrected diffraction patterns as described above.  Following the pole figure (double standard triangle color-set inset in (a)), using the MapViewer and DiffGen software of ASTAR package we found that some of the patterns at the top left region (red circle in (c)) are close to the Ni-[102] zone-axis. Within this region, \figSS{calib}(a) shows the pattern from pixel x=26, y=27, which corresponds to the Ni-[102] with an error of $1.22^\circ$. (c) Convoluted (a) and (b) maps, which reliably show the boundaries between the amorphous BMG and Ni crystalline phases.}
 \label{fig:orientation_map}
\end{figure}

Simulated DP's for all different possible orientations with an angular step of $1^{\circ}$ were generated using a Crystalline Ni phase obtained from Crystallography Open Database (COD No 9008476) using DiffGen software of ASTAR package~\cite{rauch_improving_2010, portillo_precession_2010}.
At any particular point, the  experimental pattern with a highest correlation index value against all possible orientations (here the Ni single phase) from simulated DP's provided an adequate crystal phase and orientation or each point in the map.
The estimated orientation matching indexes are plotted in \figSS{orientation_map}(a) and represented in a stereographic color projection  (reduced  to  a double  standard triangle  color set). 

Reliability indexes, which represent the goodness of fit between the experimental DP's and the simulated ones, are calculated from the best two solutions that match the experiential pattern, and mapped into \figSS{orientation_map}(b) (brighter color represents  a more reliable fit of orientation). \figSS{orientation_map}(b) shows that the amorphous BMG region have practically zero reliability to any Ni orientation, as expected. The convoluted map between the orientation (\figSS{orientation_map}(a)) and the reliability (\figSS{orientation_map}(b)) maps (shown in \figSS{orientation_map}(c)) classifies between the BMG and crystalline Ni reliably, and confirm our concluded classification from the VBF image shown in \fig{overview}(d)(i). For distortion-correction and calibration processes, we chose the DP in the pixel $x=26$, $y=27$ (\figSS{calib}) as our standard. This pixel is placed within the red square in \figSS{orientation_map}(c). The DiffGen simulation estimated a fitting error of $\pm 1.22^\circ$ to the DP.

\subsection{Refined distortion correction}  \label{sec:refined_distortion_correction}

\begin{figure}[h]
 \centering
 		\includegraphics[width=0.85\linewidth, valign=t]{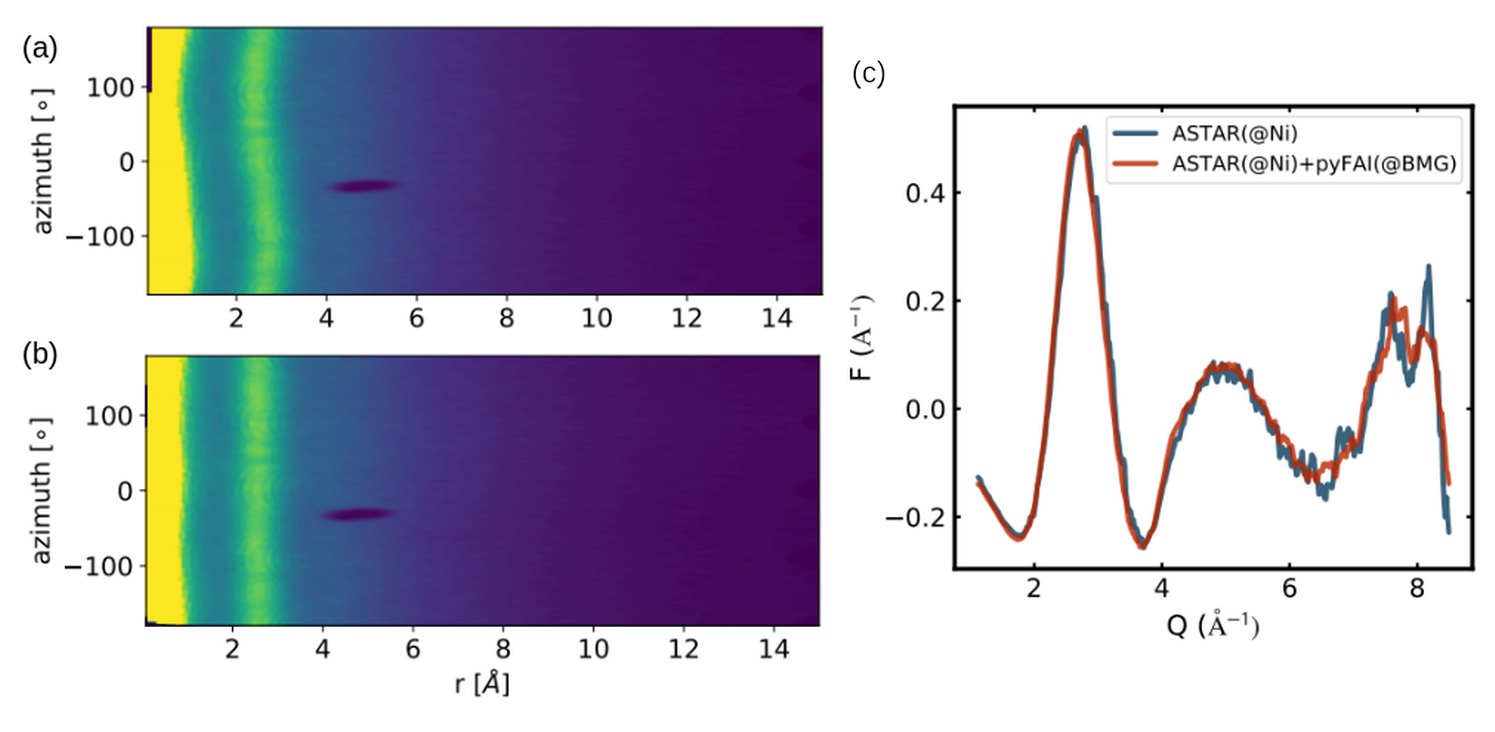}
 \caption {Comparison between distortion-correction approaches using (a) a diffraction pattern from a BMG region after distortion-correction using the ASTAR software from a crystalline Ni pattern [102] (see \sect{distortion_correction}). (b) Additional distortion correction for remaining ellipticity in the DP's using the pyFAI calibration module from an amorphous BMG pattern (see \sect{refined_distortion_correction}). (c) The azimuthally-integrated reduced structure function F(Q) function as resulted from the integration of (a) and (b), which shows that the additional correction reduced the noise level and peak width, as expected.}
 \label{fig:radial}
\end{figure}

The distortion correction carried out in the ASTAR software was found to result in Debye-Scherrer rings that were not perfectly circular and a second distortion correction was found to improve the quality of resulting PDFs signficantly. A further refinement of the parameters that define rotations of the detector away from being perpendicular to the beam were carried out in the pyFAI calibration module~\cite{kieffer_pyfai_2013} on a DP from an amorphous region of the sample. The improvement in the calibration can be seen as the unfolded Debye-Scherrer rings appearing as straight vertical lines in \figSS{radial}(b) rather than wavy lines in \figSS{radial}(a). The azimuthally-integrated DP reduced structure function, $ F(Q)$, which is the function that is directly Fourier-transformed to a PDF (see \sect{get_pdf}) shows  that the additional distortion correction reduces the SNR in the higher $Q$ range and a narrower peak-width. This additional correction allowed us to use data over a higher $Q$-range for constructing a PDF, thus improving the real-space resolution and resulted in better quantitative insight from the resulting ePDF's.

\subsection{$Q$-Calibration} \label{sec:calibration}
The 2D diffraction pattern must be integrated along arcs of constant $Q$ to obtain a 1D powder diffraction patter that is Fourier transformed to obtain the PDF. The approach is fairly standard and is described in detail in the supplementary information in \sect{calibration_details}.

We note that, in principle, after image correction and centring, the calibration is not necessary if one wishes only to get a relational comparison, meaning follow differences between different areas of the sample, rather than extracting absolute structural parameters. The PDF obtained from uncalibrated data will appear the same as from calibrated data, but the positions of the PDF peaks will not accurately reflect the real interatomic distances.  We mention this because correct calibration can sometimes be challenging, but important and valid scientific insights can still be extracted from data that have only an approximate calibration, or no calibration at all.

The basic approach for calibration is to obtain, under the same microscope conditions as the SNEM measurement, a known diffraction pattern from a reference material.
Here we took advantage of the fact that outside the BMG we are looking at crystalline nickel and it was possible to find a diffraction pattern from a crystallite down a known zone axis [102].
Other $Q$-calibration strategies can also be found here~\cite{walck_recipes_2020}.

\subsubsection{Calibration details}
\label{sec:calibration_details}
This section describes a calibration process which allows accurate 1D powder diffraction patterns to be obtained from the native 2D diffraction images on the detector.
First, image distortions are removed using the ``ASTAR"~\cite{rauch_improving_2010, portillo_precession_2010} package that corrects for any physical misorientations of the detector with respect to the incident beam, as well as other distortions coming from the experimental setup. The distortion correction is carried out such that a known single-crystal diffraction pattern appears undistorted on the detector. For example, as discussed below, in our case we used a diffraction pattern from parts of the crystalline nickel part of the sample. For more details about the distortion-correction see \sect{distortion_correction} in the main text. The resulting corrected diffraction image yields an image of a plane through reciprocal space, that sufficiently close to an image that would be obtained with an average incident beam hitting the detector normal to the plane of the detector. The ASTAR software yields a highly accurate position of the beam center on the detector. It also yields the pixel edge length in reciprocal lattice units, where the dewarping corrections result in a sufficiently linear (in reciprocal space) grid of pixels and the pixels are square.
The beam center and the reciprocal-space size of the pixels therefore provide enough information to determine the $Q$ value of every pixel in the image, which completes the calibration.

\begin{figure}[h]
 \centering
 		\includegraphics[width=0.75\linewidth, valign=t]{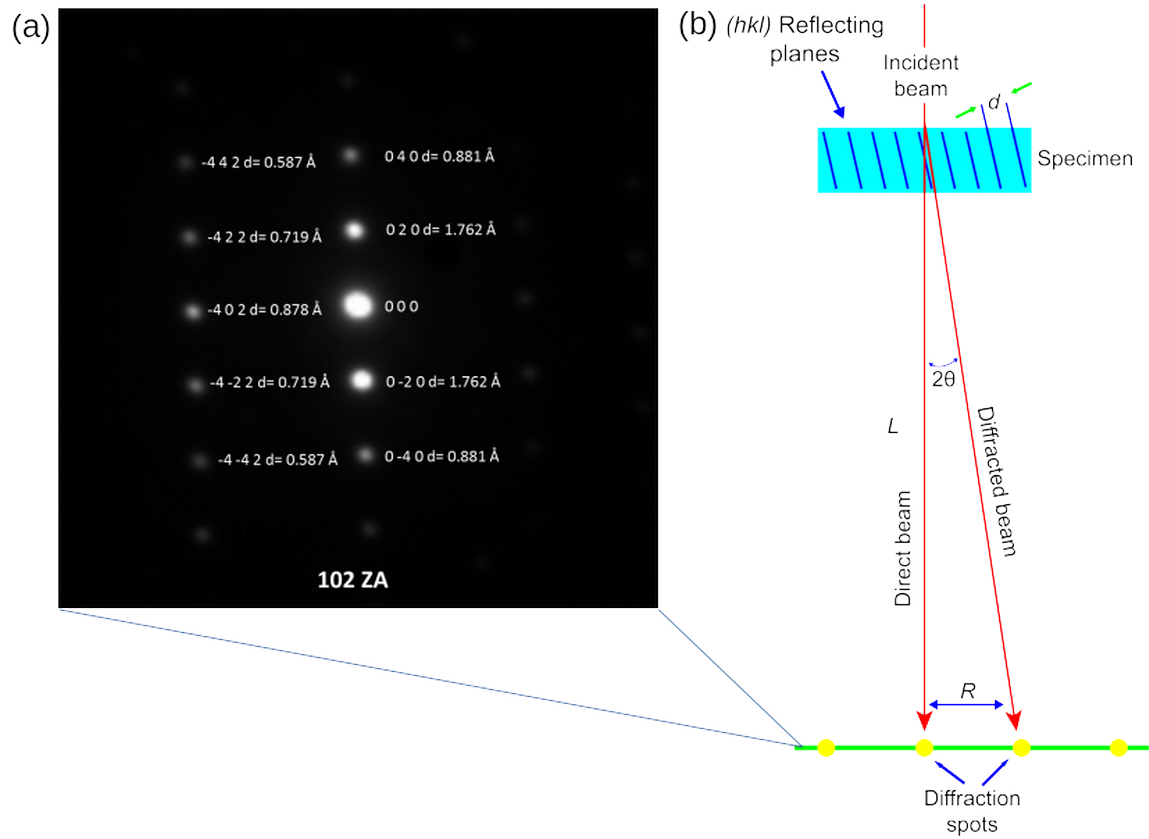}
 \caption{Calibration: (a) a calibration diffraction image for determination of the effective Q resolution (the diffraction pixel size). This image correspond to the x=26, y=27 indexed diffraction pattern (indexes relate to \fig{intro}(b)). Using the orientation matching map (\figSS{orientation_map}), this diffraction image was found most suitable to the Ni-[102] orientation.
(b) A schematic illustration of a classical (e.g., x-ray) diffraction experiment in a transmission mode, without the magnification lenses that are present in electron diffraction. This emphasizing the relevant geometries used for extracting the ``calibration parameter'', $\mathrm{\dfrac{(camera~length)}{(pixel~size)}}$. In the illustration, $L$ is the camera length, $d$ is the crystallographic spacing and $R$ is the distance between the incident beam and a diffraction spot. $R$ can be also defined as $\mathrm{(num.~of~pixels)\times (size~of~pixel)}$}
 \label{fig:calib}
\end{figure}

For the current paper we used the calibration image in \figSS{calib}(a), which is a diffraction pattern from the $x=26$, $y=27$ on the sample that corresponds to a Ni crystallite close to the [102] zone axis (actually, with an offset of $1.2^{\circ}$ from the zone axis).  The ASTAR software yielded a pixel edge dimension of $\mathrm{0.0319 \AA^{-1}}$ for the pixels and a beam center at the x=72, y=72 pixel of each dewarped DP.

To determine the 1D powder pattern, which is also needed for the PDF, it is necessary to integrate along arcs of constant-$Q$ on the detector.
In order to carry out the integration, we use software that was developed for integrating x-ray diffraction patterns from 2D detectors such as SrXplanar~\cite{yang_xpdfsuite:_2014,noauthor_srxplanar_nodate}, pyFAI~\cite{kieffer_pyfai_2013} or Fit2D~\cite{hammersley_fit2d_2016}.  Because of the particular geometry of the x-ray measurement, with scattered beams moving in straight lines from the sample to the detector, these programs do not take as input the reciprocal-space dimensions of the pixel and it is necessary to convert the information obtained from ASTAR to the relevant input for the integration program.

In the current case we used pyFAI for the integration. In this case the relevant inputs are a physical sample-to-detector distance measure (strictly the distance of normal incidence), $L$, the position on the detector of the beam center, the wavelength of the incident radiation, $\lambda$, and the physical pixel dimensions in units of meters.
pyFAI also takes detector angular offsets as inputs. The dewarping corrections result in an image that would be obtained from an incident beam with normal incidence, so all detector angular offset parameters can be set to zero in the inputs. $\lambda$ and the beam center position are known. However, it remains to determine an ``effective camera length", $L$, which is the length the camera would have had if there was no magnification or beam bending of the scattered electrons, to use pyFAI.

The geometry for the effective camera length definition is shown in \figSS{calib}(b).
We need to convert the known pixel size in reciprocal-space into an ``effective camera length", $L$.  A length in reciprocal space, $Q$, is given by
\begin{equation}Q = 4\pi\sin\theta/\lambda,\end{equation}
where $\theta = 2\theta/2$ and $2\theta$ is the angle subtended at the sample between the direct beam and the scattered beam.  From the effective camera length geometry, we see that $\tan (2\theta) = R/L$ and in the small angle approximation $\tan (2\theta)  = R/L \sim 2\theta \sim 2\sin\theta$ and $\sin\theta = R/2L$.  Thus, if we take $R$ to be the distance from the corner of one pixel to the corner of the next pixel, i.e., the physical pixel dimension in our image, $R_p$, and $Q_p$ to be the known pixel dimension in reciprocal space from the ASTAR calibration, we get that
\begin{equation} L = \frac{2\pi R_p}{Q_p \lambda}.\label{eq:ecl1}\end{equation}
In principle, the integration program wants $L$ and $R_p$ in units of meters, but provided the small angle approximation is satisfied, these two quantities will be divided by each other in the calculation and so it is possible to give these quantities in any length units as long as they are both in the same units. For example, we can give in units of ``pixel edge length", $R_p$, in which case Eq.~\ref{eq:ecl1} becomes
\begin{equation} L = \frac{2\pi}{Q_p \lambda}.\label{eq:ecl2}\end{equation}
The value of $L$ determined this way will then be used as input to the integration program as the camera length variable, with the $x$ and $y$ pixel dimension inputs set to~1 (for the current case of square pixels).
In the current experiment, using the electron wavelength for a 200~keV beam that is relativistically $\mathrm{\lambda=0.0251~\AA}$ and $Q_p=0.20$ from ASTAR, we get an effective camera length of $L=1249$ in units of pixel dimension.

We note that this conversion will work if we know the $Q$-value of any feature in our image.  For example, if we find the center of a known diffraction spot, for example, the Bragg peak from a known set of crystallographic planes with Miller indices $(hkl)$, we can determine the distance on the image from the beam center position to the center of the Bragg spot in units of the pixel edge length, $N_p$ (most simply  by counting the pixels along the $x$ and $y$ directions and using Pythagoras' theorem), in which case
\begin{equation} L = \frac{2\pi N_p}{Q_{hkl}\lambda}.\label{eq:ecl3}\end{equation}
Since $Q_{hkl}=2\pi/d_{hkl}$, where $d_{hkl}$ is the interplanar spacing of the set of planes $(hkl)$, this can be rewritten as
\begin{equation} L = \frac{d_{hkl} N_p}{\lambda}.\label{eq:ecl4}\end{equation}
This allows an approximate calibration to be carried out without a well defined $Q_p$ from the ASTAR software.  The calibration will be less accurate than using a more sophisticated regression algorithm to extract the pixel edge length, as is done in ASTAR (that is using this approach internally to do the calibration from a known diffraction pattern) but in many cases could be sufficient.

These conversions are valid when the small angle approximation is valid, or when $L \gg R$.  To estimate the largest error introduced by taking this approximation we consider the largest $R=R_{max}$, which is the distance from the beam center to the corner of the detector.  The largest error, $\delta Q_{max} = 4\pi(\sin\theta_{max} - \theta_{max})/\lambda$.  In our experiment having the beam center placed at the center of $144\times 144$ pixels, $R_{max}=101.8 pix$, and having $L=1249$ pixels, form $tan(2\theta_{max})=R_{max}/L$, we get that $\theta_{max} = 0.041~rad$ and $\delta Q_{max} = 0.003$.

\section{SNEM data analysis} \label{sec:data_analysis}
\begin{figure}[h]
 \centering
 		\includegraphics[width=0.95\linewidth, valign=t]{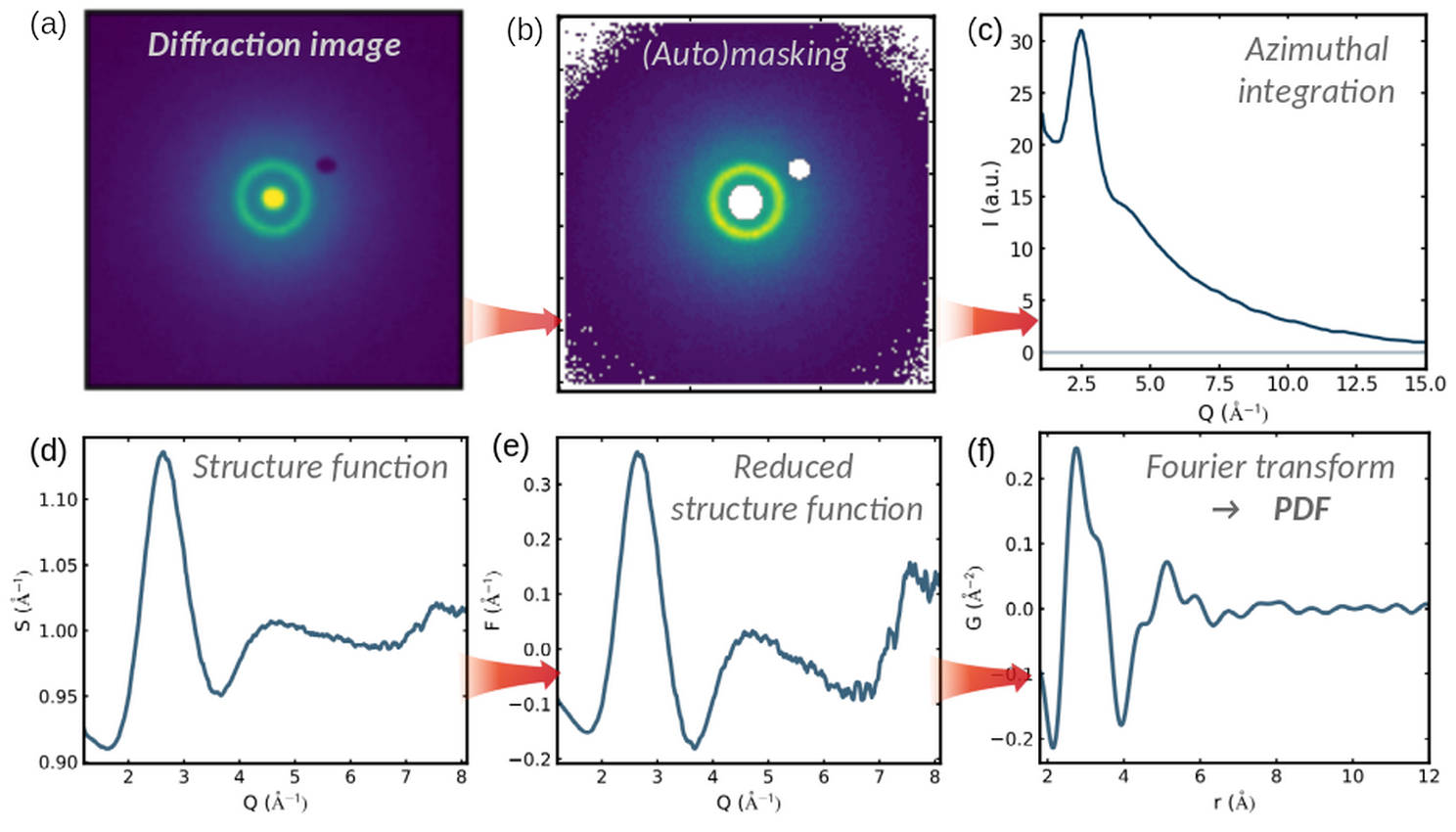}
 \caption {The analysis pipeline that was used to generate a single PDF from a DP. (a) Raw diffraction image; (b) masked diffraction pattern to remove unwanted pixels (central beam, over-saturated or bad pixels, isolated crystalline diffraction peaks, etc.) (c) azimuthally integrated intensity of the masked image, $I(Q)$; (d) structure function, $S(Q)$, (\eq{sq}); (e) reduced-structure function, $F(Q)$, (\eq{fq}); (f) PDF, $G(r)$, from a sine Fourier transformation of $F(Q)$. The DP in (a) is an example taken from a position $x=140$ and $y=70$ within the BMG and for the Fourier transformation we used $Q_{max}=8.1$. Other examples for data-reduction of DP's from other regions where auto-masking was required to remove Ni peaks closer to the Ni/BMG region are shown in \figSS{automask}}
 \label{fig:point}
\end{figure}

After distortion correction, centering and calibration we can initiate the data reduction pipeline towards generation of PDF's from each individual diffraction pattern. This will allow us to expose and see more vividly local structural features. \figSS{point} represents the steps in the data reduction pipeline from a distortion corrected image in panel (a) to the PDF  in panel (f). In the following subsections we describe the different steps.

\subsection{Auto-masking} \label{sec:get_automasked}
\begin{figure}[h]
 \centering
 		\includegraphics[width=0.80\linewidth, valign=t]{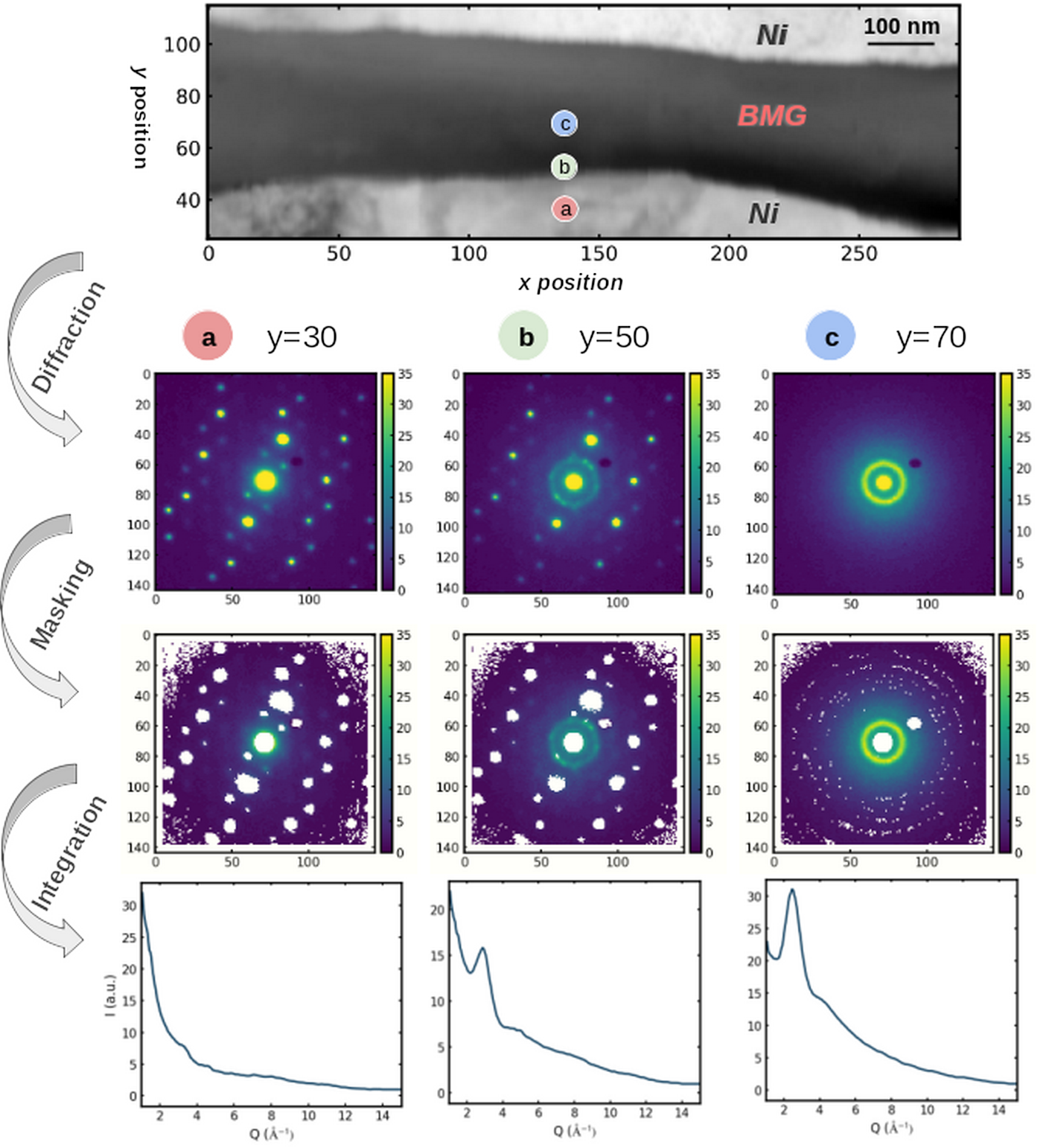}
 \caption{Examples of three characteristic points along a cut x=140, where the points a,b and c represent the estimated position of where  y=30, y=50 and y=70, respectively. The top most image is the VBF with addition of the a, b and c points. For each point there is a representation of the raw diffraction image, the masked image, the integrated $I(Q)$ plots.}
 \label{fig:automask}
\end{figure}

Our laminate sample contains two distinct regions:  crystalline nickel and amorphous BMG. The diffraction signal from regions containing crystals have bright diffraction spots as shown in \fig{overview}(b) and \figSS{automask}. \rev{1.2}{Since conventional PDF analysis require isotropic diffraction patterns, we wish to remove the signal from the crystalline component and focus on the diffuse signal that contains the local structural information of the BMG. To do this we use an auto-masking algorithm, where we automatically determine the location of Bragg peaks and mask their intensities from those points, removing them from each diffraction pattern.}

First, a static mask was applied to remove the direct beam spot and dead pixels that always appear at the same spot on the detector. Then, to remove Bragg peaks,  (and any other occasionally appearing dead pixels) an auto-generated mask was applied. The auto-generated mask is generated using the `xpdtools' software package~\cite{xpdtools}. The mask auto-generation algorithm  iteratively finds and removes outlier pixels around rings of constant radius~\cite{xpdtools, wright_computer-assisted_2017}. Since a truly amorphous signal has a normal distribution of intensities along an azimuth with a constant radius, the algorithm detects intensities from crystalline particles, which then masked.  This process allows us to isolate the contribution from the BMG, as demonstrated in \figSS{automask}. The auto-masking configuration parameters we used for this work are summarized in \tablSS{params}.

\rev{1.2}{We note that the auto-masking algorithm was used only as part of the data-reduction pipeline during extraction of QoI's from ePDF data. It was not applied, however, when extracting QoI's directly from the diffraction patterns.}

\subsection{Integration of masked diffraction images} \label{sec:get_iq}
The next step to get the signal for PDF analysis is to carry out an azimuthal integration around circles of constant momentum transfer $Q$ in the diffraction pattern~\cite{egami2012underneath}. For this we use the Python-based fast azimuthal integration package pyFAI~\cite{kieffer_pyfai_2013}, which yields a 1-dimensional intensity, $I(Q)$, as shown in \figSS{point}(c). As mentioned earlier, the $Q$ values in the $I(Q)$ and the consequent PDF obtained from uncalibrated data will appear the same as from calibrated data, but the positions of the PDF peaks will not accurately reflect the real interatomic distances. Here, following \sect{calibration}, we used a diffraction pattern from a reference material to determine the accurate $Q$ spacing in each DP.

\subsection{Data reduction to obtain the PDF} \label{sec:get_pdf}
The next steps are to take the 1D intensity function, correct and normalize it to obtain the reduced total scattering function, $F(Q)$ that can be Fourier transformed to obtain the PDF. Here, $F(Q)$ is defined as~\cite{egami2012underneath}
\begin{equation}
F(Q)=Q[S(Q)-1],
\label{eq:fq}
\end{equation}
where
\begin{equation}
S(Q)=\frac{I_c(Q)-\langle f(Q)^2\rangle + \langle f(Q)\rangle ^2}{\langle f(Q)\rangle ^2}
\label{eq:sq}
\end{equation}
is the reduced structure function. $I_c(Q)$ is the intensity from coherent scattering events (corresponds to the azimuthally integrated data from the prior step) and $f(Q)$ is the electron atomic form factor~\cite{abeykoon_quantitative_2012}. $F(Q)$ was then Fourier-transformed to $G(r)$, which is the PDF:
\begin{equation}
G(r)=\frac{2}{\pi}\int_{Q_{max}}^{Q_{min}}F(Q)\sin(Qr)dr.
 \label{eq:gr}
\end{equation}
For these steps we used the program PDFgetE3, which is part of the ePDFsuite~\cite{epdfsuite} software.

\section{Definition of Quantities of Interest} \label{sec:get_qoi}
\begin{figure}[h]
 \centering
 \includegraphics[width=0.95\linewidth, valign=t]{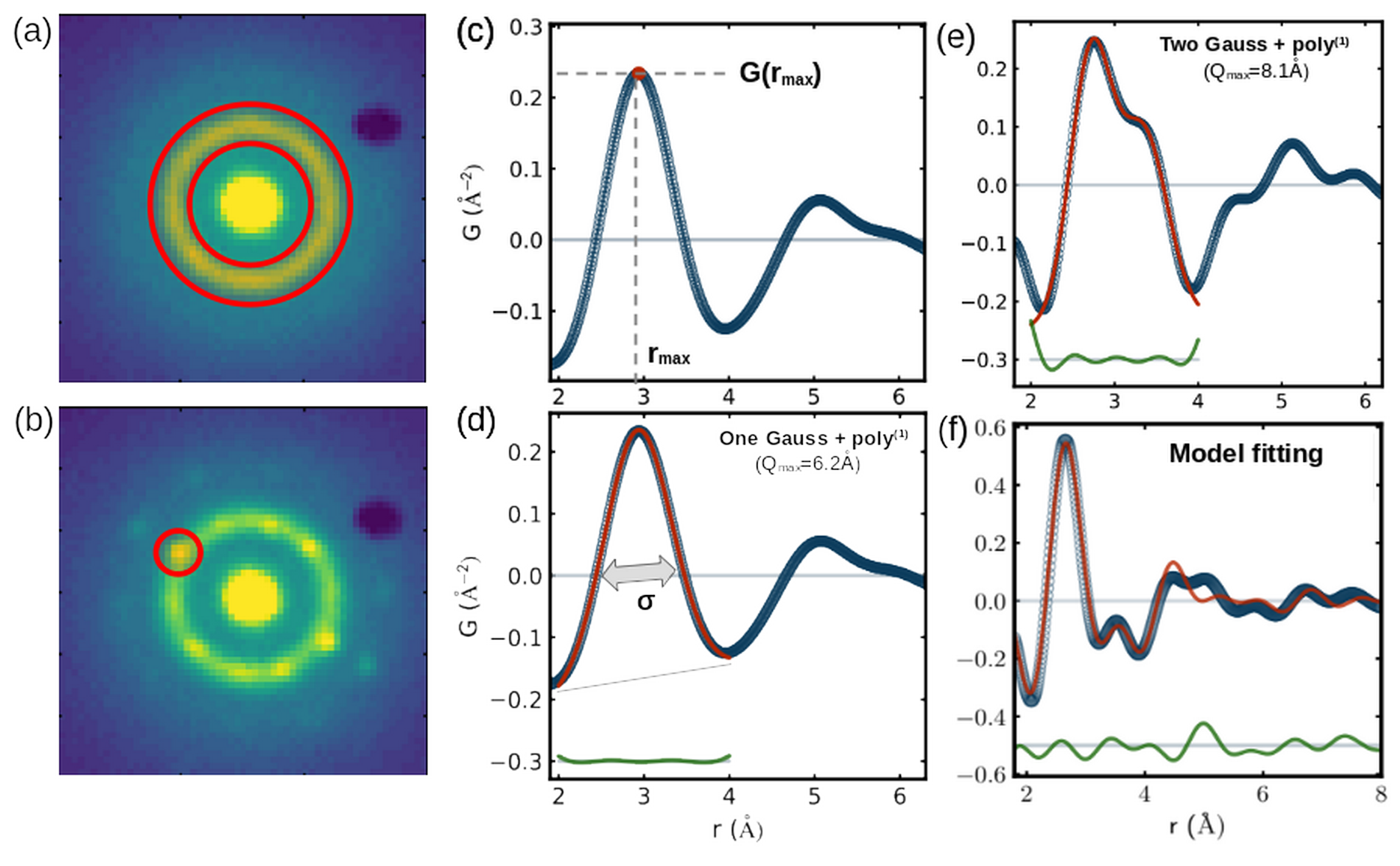}
 \caption{Visualized examples of QoI value extraction from each 2D diffraction pattern and 1D-ePDF. (a) and (b) are RoI's within a 2D pattern from which the \textit{sum} or the \textit{max} intensity can be derived as a QoI; while (a) is agnostic to a particular orientation and gives an overview over all azimuthal orientations, (b) is specific to a specific orientation. (c) Extraction of the average maximum coordinates QoI's, $r_{max}$ and $G(r_{max})$, from the nearest-neighbouring peak. (d) and (e) shows extraction of QoI's from a fit to Gaussians on top of a firs-order polynomial (\eq{gauss}) from which the peak width ($\sigma$) and maximum can be extracted as QoI's. For the example of (d) we use $Q_{max}=6.2$ to illustrate a PDF fitting to a single Gaussian. In this work, however, we use the highest possible $Q_{max}=8.1$\AA$^{-1}$ whcih produce a double Gaussian to which we fit \eq{gauss} with N=2, as demonstrated in (e). (f) represents a more complex fitting of a PDF to a structural model from which each fitted parameter is a potential structurally-significant QoI. Throughout the paper, and as show in panels (d)-(f), the red line refers to the fit or model data, the blue circles represent the data and the green green line to the difference between the data and the fit. Initial fitting and boundaries conditions are summarized in \tablSS{params} (for (d) and (e)) and \tablSS{fit_params} (for (f))}
 \label{fig:qois}
\end{figure}

Any scalar quantity (number) that can be extracted from the data can be a ``quantity of interest" (QoI), if by mapping its value as a function of position across the sample we glean some information about the local structure or chemistry.  \figSS{qois} summarizes a number of QoIs we extracted in the current study.

%

Powerful and widely used QoIs from the raw data are virtual apertures that define diffraction image regions of interest on the detector, for example, circular RoI apertures may be used to produce bright and dark field images, sometimes called ``virtual" bright and dark field (VBF, VDF) images~\cite{rauch_virtual_2014, ophus_four-dimensional_2019}. For a VBF map, the circular diffraction RoI is centered on the main beam.  In a conventional bright field image it is proportional to the transmitted beam and the contrast in the map represents the thickness and scattering power of the sample from place to place. In our virtual bright field images, detector saturation by the main beam was found to distort this image with counts suppressed in the most highly saturated pixels.  We therefore built masks to remove the heavily saturated center part of the beam.  As with conventional dark field imaging, circular apertures may also be defined to select circular diffraction RoIs on any interesting feature in the 2D diffraction image, and the resulting spatial map is a virtual dark field image of the phase producing this diffraction feature. For example, the contrast in a VDF map can expose the distribution of crystallographic orientations of a known phase~\cite{ophus_four-dimensional_2019} (\figSS{orientation_map}(a)), and in our current study, the presence of nanocrystallites in the BMG (\figSS{qois}(b)).

However, the QoI concept allows us to define many more potentially interesting RoI's. For example, by defining an RoI that is a circular annulus that captures the strongest diffraction ring of the BMG, (\figSS{qois}(a)) and summing the intensity in all the pixels that fall in that annulus, we can map out the location of the BMG in the image (\fig{overview}(d)(i)).

After reducing the diffraction patterns into PDF's, we can then define QoI's from features in the PDF function that are directly giving information about inter-atomic distances in the local structure.
For example, the first peak in a PDF comes from the distribution of pair-distances within the first coordination shell. Since the BMG structure contains atoms of different sizes, measuring the position of the first-peak maximum, $r_{max}$, should reflect the (scattering weight averaged) average pair-distances within the first coordination shell, and is therefore tightly related to the compositional distribution.  The peak needs to be modelled to get this information in a highly quantitative way since in a mixed system the average peak position will be weighted by the scattering power of the atoms involved as well as the interatomic pair lengths.  However, simply mapping $r_{max}$ (the $r$-position of the PDF peak maximum) vs. position can give an indication of compositional variations in the sample, very rapidly and easily.  Other QoI's may also be plotted, such as the value of $G$ at its maximum, $G(r_{max})$.  The meaning of this QoI is less clear because PDF peak height depends on the sharpness of the peak as well as the  scattering strength (and therefore composition), but QoI maps from this QoI can still give valuable information about compositional (peak height) and structural (peak width) variations. A schematic of how these two QoIs are defined, is shown in \figSS{qois}(c).

We can go further by fitting models to the PDFs and using refined model parameters as our QoIs and mapping how they vary spatially.
For example, to separate the contributions of sharpening and integrated intensity to the first PDF peak maximum we can fit one or more Gaussian functions to the first PDF peak, as shown in \figSS{qois}(d). We find these fits more satisfactory when we add a linear baseline.  The fit results in three parameters for the Gaussian, its center, width and integrated intensity, and these three values can be independently mapped.  In general, the first peak in the PDF may be made up of multiple interatomic distances, and requires a fit with a sum of Gaussians,
\begin{equation}
	G^{Gauss}(r) = \sum_{i=1}^{N}\left( \frac{A_i}{\sqrt{2\pi} \sigma_i} \exp^{\left(-\frac{(r-r_{i(cen)})^2}{2\sigma_{i}^2}\right)} \right) + P^{(1)}(r)
 \label{eq:gauss}
\end{equation}
where $A$ is the integrated area, $r_{cen}$ the center of the Gaussian, and $\sigma$ is the standard deviation (peak width). $P^{(1)}(r)$ is a first order polynomial function ($P^{(1)}(r)=a\cdot r+a_0$) that acts as a baseline for the Gaussian. The $i$ index runs over the $N$ Gaussians used in the fit.
Examples of successful single Gaussian and two-Gaussian fits are shown in \figSS{qois}(d) and (e), respectively.  The fit range must also be defined.
When fitting \eq{gauss} we used scipy.optimize.curve\_fit) from the python `scipy' package~\cite{2020SciPy-NMeth}. The initial fitting parameters are found in \tablSS{params}.

Alternatively, one can come up with a small box structural mode and fit it to a PDF using well established softwares, such as PDFgui~\cite{farrow_pdffit2_2007} or diffpy-cmi~\cite{juhas_complex_2015} for more complex cases. The fitting parameters, including the degree of agreement between the model and the structure can be then used as a reasonable QoI. This process is thought cumbersome and require careful examination of the fitted model. If done successfully with all the cautionary required, the richness of the structural information  is large, for example, the explicit interatomic bond-distances, bond densities the agreement with a structure with respect to a competing structure, etc. 

In this work we assumed a model of a closed-packed FCC structure that contains three types of bonds: Zr--Zr, X--X and Zr--X. To find the partial contribution for each bond we split the structure into three FCC sub-structures of Zr--Zr, X--X and Zr--X. For Zr--X we allow counting the contribution of only the Zr--X pairs and neglect any contribution that comes from Zr--Zr or X--X pairs. Using the scaling factor values, we extracted the occupancy of each pair, or in other words, the pair density. For the fitting we used PDFgui~\cite{farrow_pdffit2_2007}, where the exact fitting conditions are summarized in \tablSS{fit_params}

\section{Software} \label{sec:software}
Besides data acquisition, calibration and distortion correction, the analysis pipeline strives to be Python-based. The analysis pipeline software that are mentioned in the Methods section, including for masking, azimuthal integration, Fourier transformation, QoI generation and imaging, are wrapped by a Python-based graphical user interface, called `MiniPipes'~\cite{rakita_genesis-efrcminipipes_2021}. MiniPipes focuses on the ease in configuring the different steps and the capability of chaining the different analysis building-blocks into a coherent pipeline that once properly configured using the user interface, can be used to rerun data analyses for convenient exploration of the data, with different QoI's. Moreover, MiniPipes allows an easy setup of new pipeline building-blocks, which allows easy implementation of new analysis pipelines and improved SNEM QoI's.

\section{EELS imaging} \label{sec:eels_imaging}
For comparison with the PDF-derived QoI maps compositional information about the sample was obtained from EELS measurements. STEM-EELS data were acquired separately, but from the same area from which the SNEM patterns were taken. A Gatan Imaging Filter Quantum was used with a K2 direct electron detector. The STEM convergence semi-angle was 8~mrad and the collection semi-angle was 45~mrad. Elemental composition maps were generated using a power-law background subtraction and Hartree-Slater cross-sections.


\newpage
\clearpage
\begin{table}[h]
\large
  \captionsetup{singlelinecheck=false, font=normalsize, labelsep=space}
\caption{The parameters used for data-reduction and QoI generation. From top: (i) automasking (for xpdtools.tools.mask\_img) parameters; (ii) data reduction and Fourier transformation (for \textit{pdfgete}/ePDFsuite) parameters; (iii) is the starting boundary fitting conditions to \eq{gauss} ; (iv) global fitting setup (scipy.optimize.curve\_fit). These parameters where applied to the entire set of diffraction images.}
\begin{tabular}{lcc}
\hline	
(i) automasking config. & description & values  \\
\hline
 auto-type 		& statistical masking strategy (mean/median)	&mean  \\	
 alpha 			& number of STD that are not masked				& 3.5    	  	\\
 edge			& number of pixels to cut from the edge 		& 3		  	\\
 lower thresh 	& lower threshold to mask 						& 1		  	\\
 upper thresh   & upper threshold to mask 						& inf 	 	\\
\end{tabular}
\\
\begin{tabular}{lcc}
\\
\hline	
(ii) data-reduction	config.  & description & values  \\
\hline
 $q_{maxinst}$ 	& Q cutoff for meaningful input intensities	&	12.0     	\\	
 $q_{max}$ 	& The lower Q-limit for the FT of the F(Q)	&	    8.1     	\\
 $q_{min}$		& The upper Q-limit for the FT of the F(Q)	&	0.1	   	\\
 $r_{poly}$ 	& r-limit for the F(Q) correction polynomial&	1.2	  	\\
\end{tabular}
\\
\begin{tabular}{lccc}
\\
\hline
(iii) fitting config. for \eq{gauss}	& & &  \\
\hline
r range [\AA]  		& 2.1 - 4 				& 						&  					\\
					& 	initial conditions	& 	lower bound			& 	upper bound   	\\
$A^{(1)}$  			&	0.01				& 	0.01	 			& 	inf				\\
$r^{(1)}_{cen}$     &	2.4					& 	2.4 				& 	3.0				\\
$\sigma^{(1)}$		&	0.01 				& 	0.01	    		& 	1    			\\
$A^{(2)}$  			&	0.01				& 	0.01	 			& 	inf				\\
$r^{(2)}_{cen}$     &	3.5					& 	3.0 				& 	4.0				\\
$\sigma^{(2)}$      &	0.01 				& 	0.01	    		& 	1				\\
$a$					&	1 					& 	-inf				& 	inf				\\
$a_0$				& 	1					&	-inf				&	inf  			\\
\end{tabular}
\\
\begin{tabular}{lc}
\\
\hline
(iv) global scipy.optimize.curve\_fit setup    \\
\hline
f\_sclae 			& 0.005 			  \\
Method  			& trf 			  \\
Loss  				& soft\_l1  	  \\
\end{tabular}
\label{table:params}
\end{table}

%
%

\newpage
\clearpage
\begin{table}[h]
\large
  \captionsetup{singlelinecheck=false, font=normalsize, labelsep=space}
\caption{The parameters used in PDFgui~\cite{farrow_pdffit2_2007} for simulation and fitting of ePDFs to a closed-packed (FCC) X--X, Zr--Zr and Zr-X structural model. The lattice parameters are calculated as $[(NN-distance) + 0.1\AA]\cdot\sqrt{2}$}
\begin{tabular}{lcc}
\hline
 parameter 					& initial value 	& free  				\\	
\hline
 scaling factor: X--X  				& 		1.0		  	&  YES   	  	\\
 scaling factor: Zr--Zr 			& 		1.0		  	&  YES		  	\\
 scaling factor: Zr--X  			& 		1.0		  	&  YES	 	  	\\
 lattice parameter: X--X [\AA]  	& 		3.68		&  YES   	  	\\
 lattice parameter: Zr--Zr [\AA] 	& 		4.67		&  YES		  	\\
 lattice parameter: Zr--X [\AA]  	& 		4.17		&  YES	 	  	\\
 ADP: X  							& 		0.005		&  no   	  	\\
 ADP: Zr 							& 		0.005		&  no		  	\\
 spdiameter [\AA] 					& 		9.0 		&  no	 	  	\\
 qdump	  							& 		0.4			&  no	 	  	\\
 fitting range [\AA]				& 	    1.8-8		&  no			\\
\end{tabular}
\label{table:fit_params}
\end{table}



%
\newpage
\clearpage
\begin{figure}[h]
 \centering
 		\includegraphics[width=0.85\linewidth, valign=t]{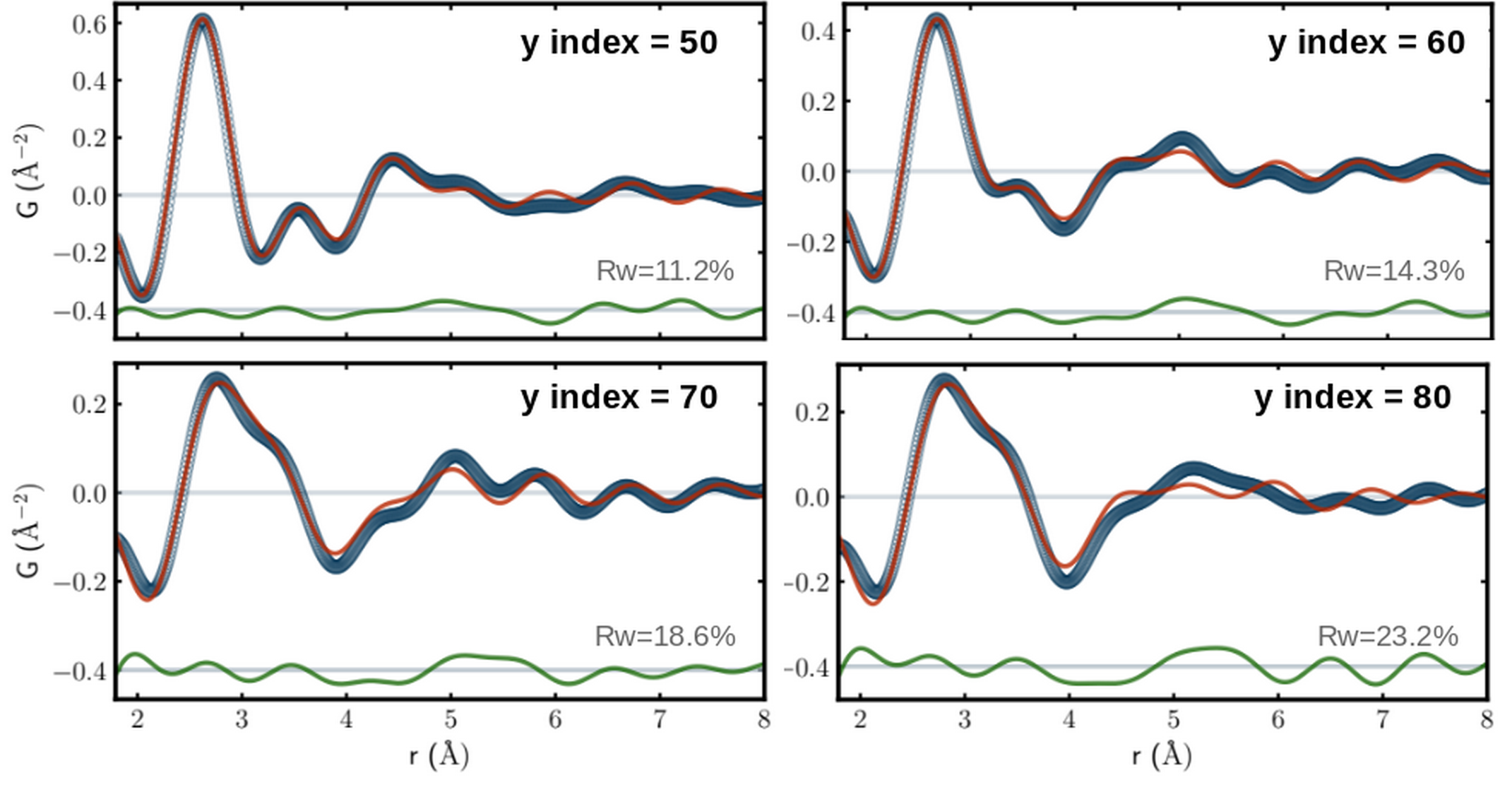}
 \caption{Fits of selected ePDFs from the x=75 cut to a superposition of X--X, Zr--Zr and Zr--X closed packed FCC structure. The y index is mentioned within each panel, where the partial PDFs of each model are plotted in \figSS{partial_pdf}. For the fittings we used PDFgui \cite{farrow_pdffit2_2007} where the fitting parameters being summarized in \tablSS{fit_params}. The fitting curve and the difference curve between the the fit and the data are plotted in red and green, respectively. The goodness of fit parameter $R_w$ ($R_w =\frac{\sum_{i}(G^{data}_i-G^{fit}_i)^2}{\sum_{i}(G^{fit})^2}$) is mentioned above the difference (green) curve.}
 \label{fig:pdffit}
\end{figure}

\newpage
\clearpage
\begin{figure}[h]
 \centering
 		\includegraphics[width=0.6\linewidth, valign=t]{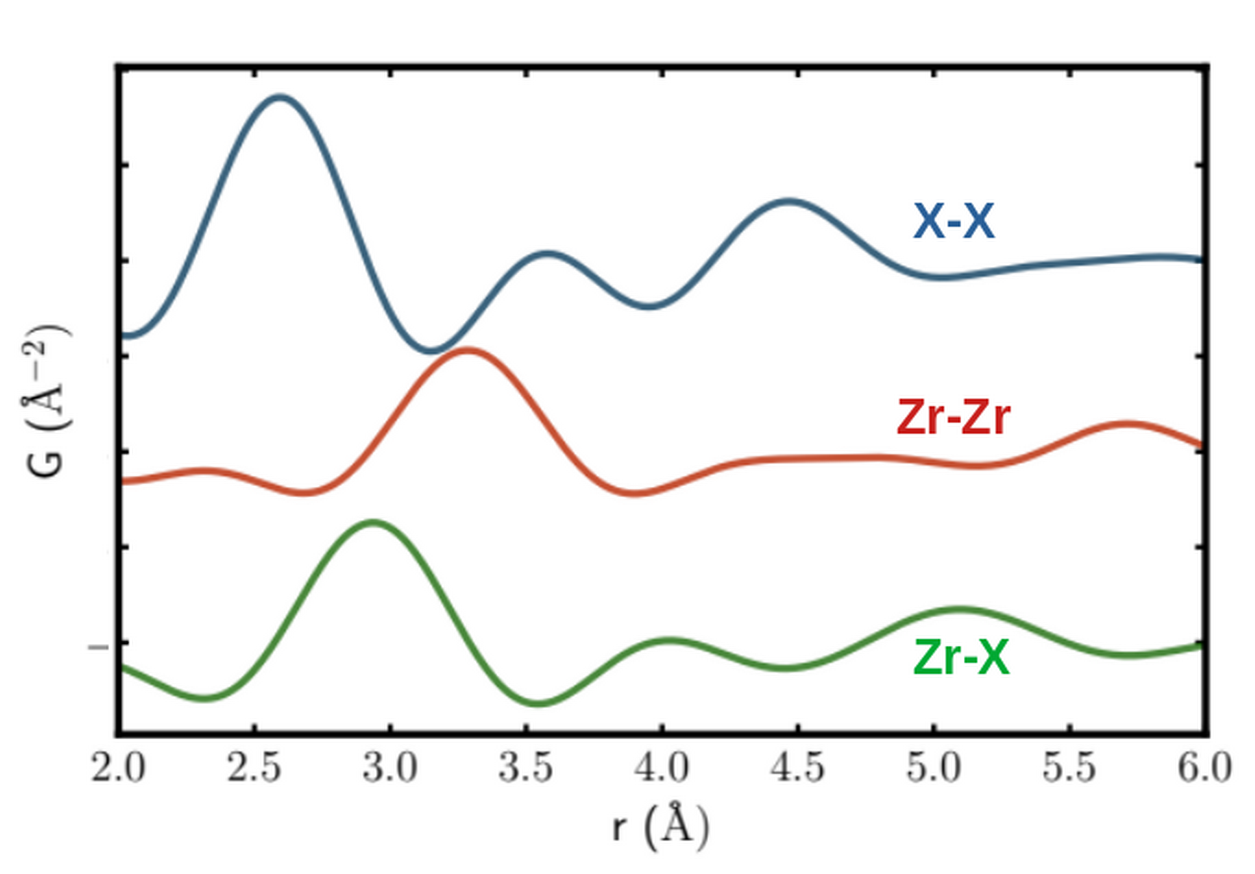}
 \caption{ Simulated partial PDFs for a closed packed (FCC) X--X, Zr--Zr and Zr--X structures. For the simulation we used PDFgui \cite{farrow_pdffit2_2007} where the simulation parameters being summarized in \tablSS{fit_params} with the exception of setting the scaling factor to 1 for the mentioned pair and 0 for the rest.}
 \label{fig:partial_pdf}
\end{figure}

\newpage
\clearpage
\begin{figure}[h]
	\centering
 	\begin{subfigure}[t]{0.80\linewidth}
 		\includegraphics[width=\linewidth, valign=t]{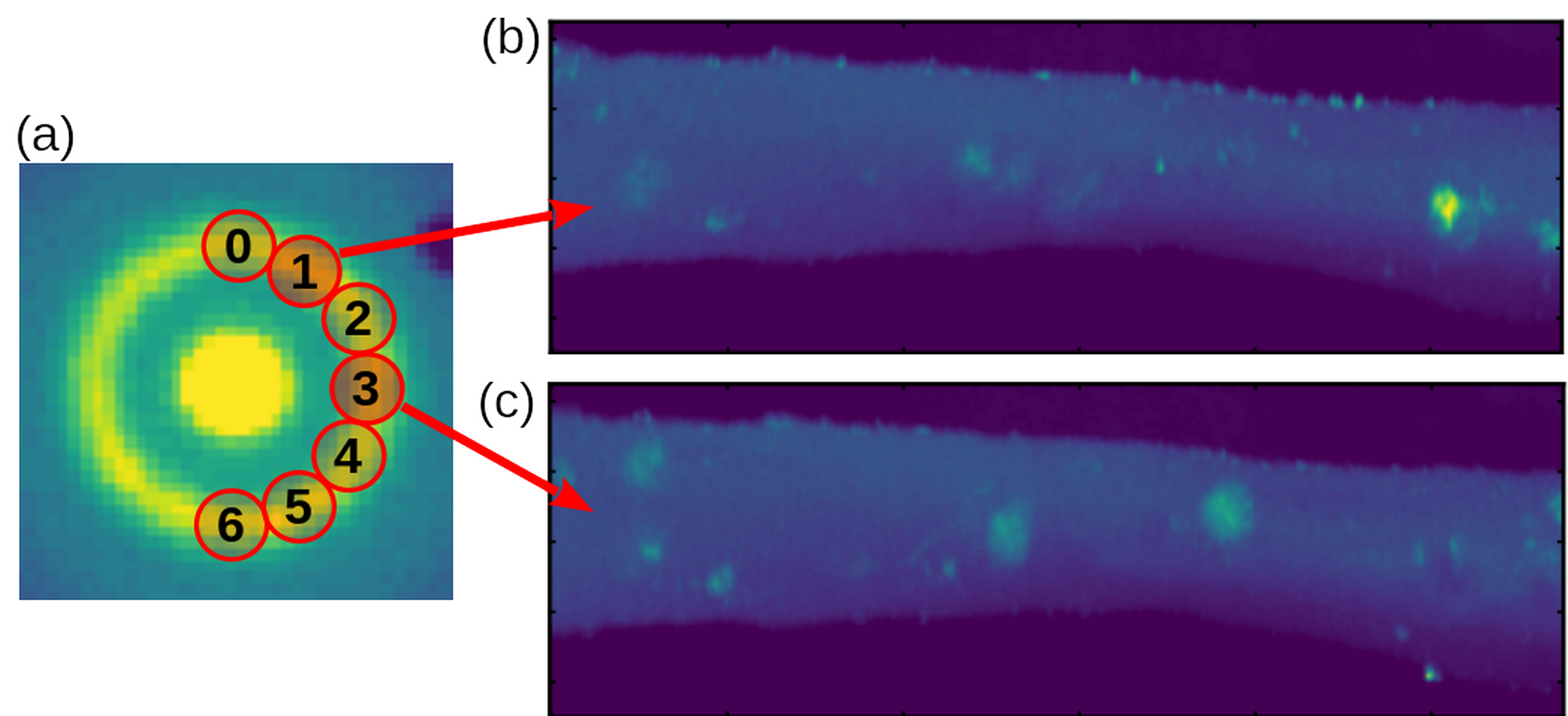}
	\end{subfigure}
 \caption{Azimuthally-selective VDF of the maximum intensity within each ROI of of a diffraction pattern. (b) and (c) are example images that are selected from region 1 and 3 in (a). A different ROI location around the ring result in a different contrast map, which relate to the crystallographic orientations  of the nano-crystalline inclusions.}
 \label{fig:VDF_point}
\end{figure}
\newpage
\clearpage
\begin{figure}[h]
	\centering
 		\includegraphics[width=0.7\linewidth, valign=t]{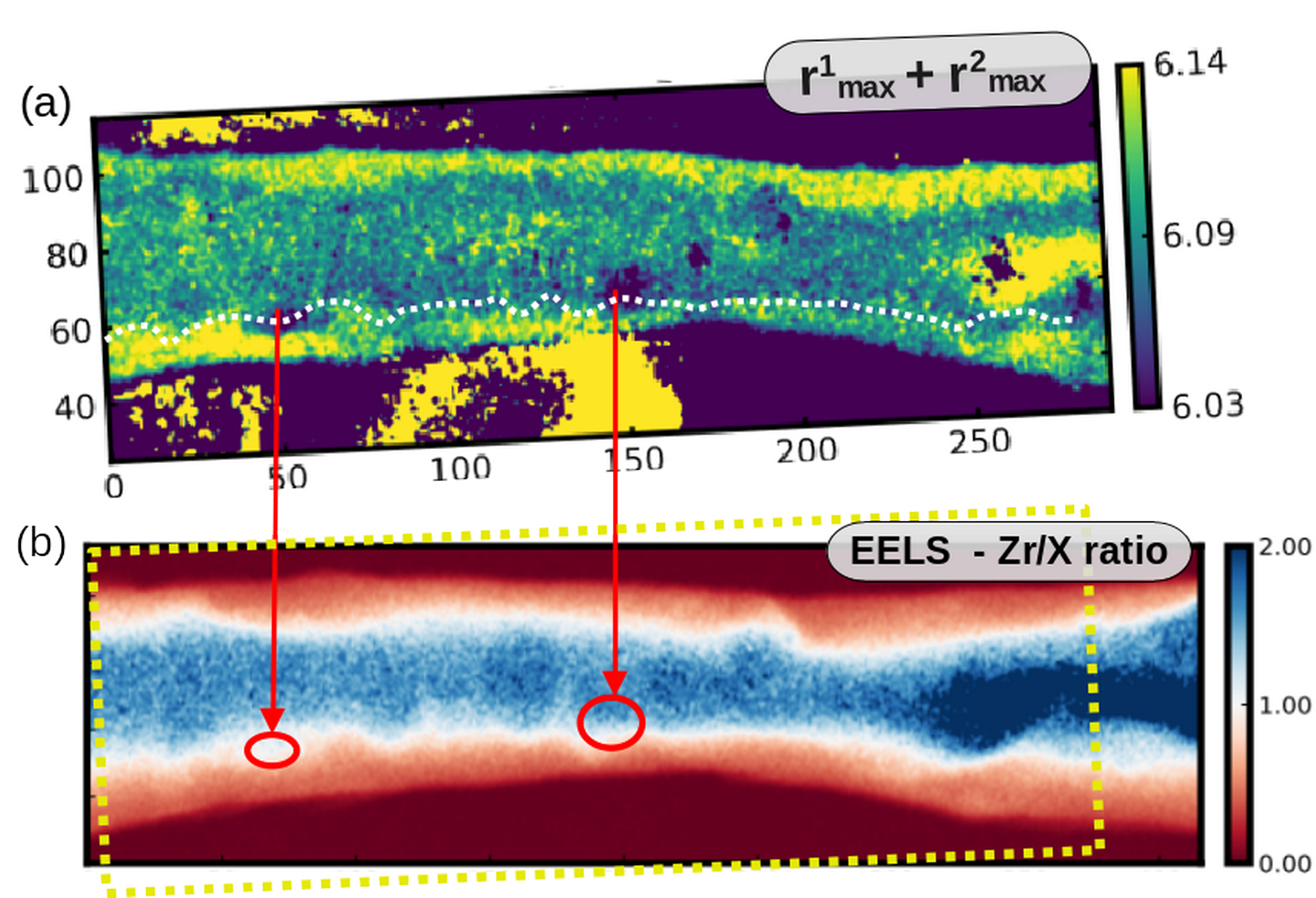}
 \caption{(a) The QoI map of $r^{1}_{max}+r^{2}_{max}$ with emphasizing the regions that suggest chemical clustering. (b) Zr/X chemical ratio derived from EELS composition mapping (\fig{eels}) showing that the Zr concentration with respect to the other elements changes mostly vertically from edge to center of the BMG and does not follow features that are shown in  $r^{1}_{max}+r^{2}_{max}$ map. Such evaluation and, especially the fact that the apparent dark regions in (a) are around [Zr--Zr] = [X--X] = 1, which are the simulation conditions for \fig{structural}(c)[blue curve], strengthen the connection between the the  $r^{1}_{max}+r^{2}_{max}$ QoI and the chemical SRO.}
 \label{fig:r1r2eels}
\end{figure}
\newpage
\clearpage
\begin{figure}[h]
	\centering
 		\includegraphics[width=0.7\linewidth, valign=t]{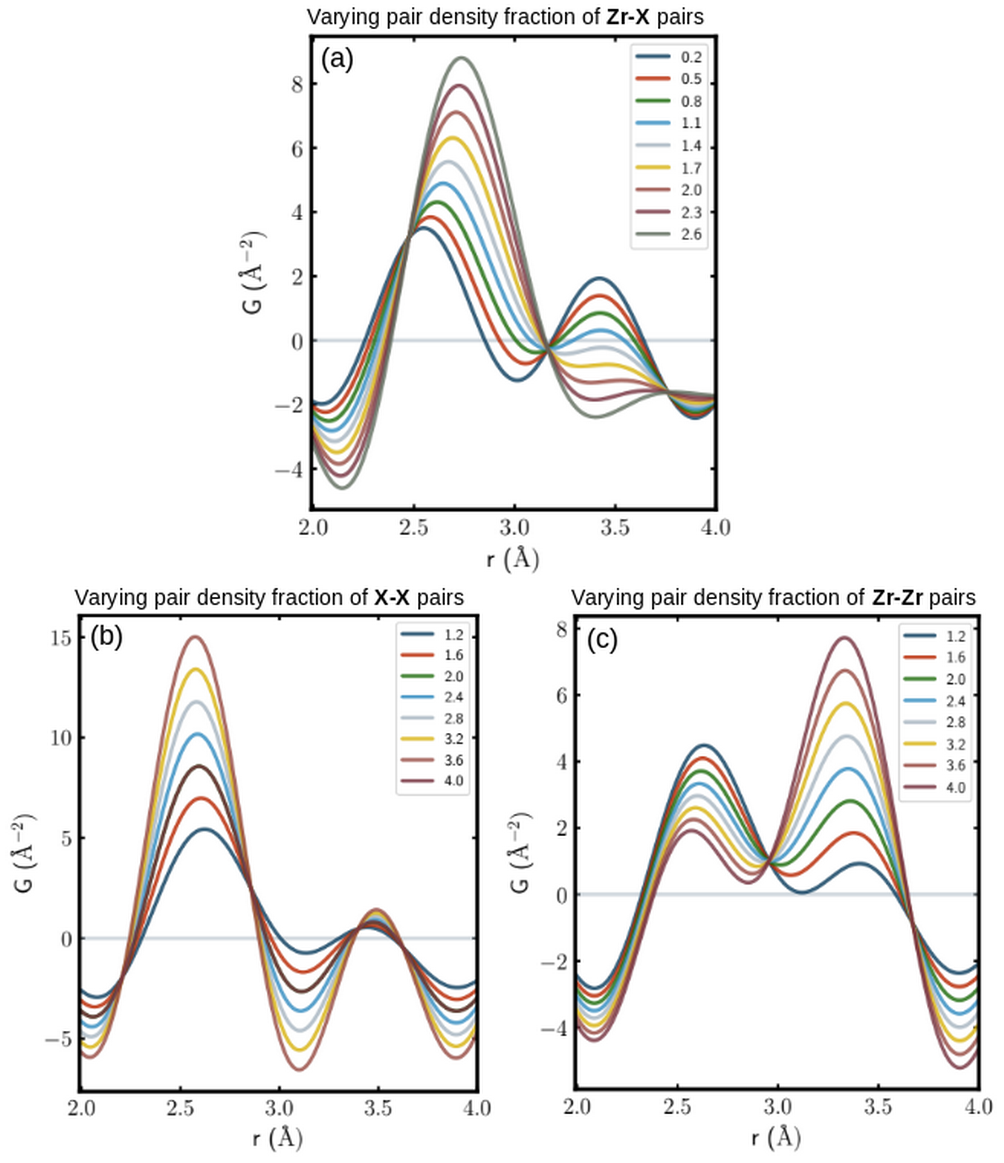}
 \caption{(a) Simulated PDFs of a closed-packed set of Zr--Zr, X--X and Zr--X structures from which $r^{1}_{max}$ and  $r^{2}_{max}$ are derived to calculate the  $r^{1}_{max}+r^{2}_{max}$ QoI for \fig{structural}(b). In (a) we kept Zr--Zr and X--X pair fraction constant and equal to 1, while changing the Zr--X pair fraction. In (b) and (c) we change the X--X and Zr--Zr fractions, respectively, while keeping the other set of pairs equal to 1.}
 \label{fig:r1r2sim}
\end{figure}
%

\label{sec:SI}

\end{document}